\documentclass{aastex}
\usepackage{spr-astr-addons}

\RequirePackage{color}

\begin{document}

\title{Orbits of Six Late-type Active-Chromosphere Binaries}
\slugcomment{Not to appear in Nonlearned J., 45.}
\shorttitle{Orbits of Six Late-type Active-Chromosphere Binaries}
\shortauthors{Griffin and Filiz Ak}

\author{R. F. Griffin\altaffilmark{1}} 
\affil{The Observatories, Madingley Road, Cambridge CB3 0HA, England \\e-mail: rfg@ast.cam.ac.uk}
\and \author{N. Filiz Ak\altaffilmark{2,3}}
\affil{Erciyes University, Faculty of Sciences and Arts, Department of Astronomy and Sapce Sciences, 38039 Kayseri, Turkey }
\affil{Department of Astronomy and Astrophysics, The Pennsylvania State University, 525 Davey Laboratory, University Park,
PA 16802 \\ e-mail: nfak@erciyes.edu.tr}

\section*{}
\begin{abstract}We present spectroscopic orbits for the active stars HD\,82159 (GS\,Leo), HD\,89959, BD\,$+39\arcdeg$\,2587 (a visual companion to HD\,112733), HD\,138157\break (OX~Ser), HD\,143705, and HD\,160934. This paper is a sequel to one published in this journal in 2006, with similar avowed intention, by \citeauthor{galvez06}. They showed only graphs, and gave no data, and no orbital elements apart from the periods (only two of which were correct) and in some cases the eccentricities. Here we provide full information and reliable orbital elements for all the stars apart from HD\,160934, which has not completed a cycle since it was first observed for radial velocity.
\end{abstract}



\section{Introduction}

In a paper with a title somewhat analogous to ours, \citet{galvez06} (cited as GMFL in what follows) presented graphs of orbital solutions for the same six stars as we discuss here. We did not set out to follow those authors, but the stars (with one exception, HD\,160934) came to our attention by featuring in the third edition of \textit{A catalogue of chromospherically active binary stars} \citep{eker08} (hereinafter called 'CABS3'), of which one of us is a co-author. They were among the minority of entries that lacked proper orbital solutions, and the other of the present authors undertook to make good some of what was missing. Twenty orbits for other such `CABS3' stars have already been presented \citep{griffin09}, and it was thought useful to discuss the six GMFL stars all together now. Basic data about them are set out in Table~\ref{table-1}.

\begin{table*}[t]
\setlength{\doublerulesep}{\arrayrulewidth}
\caption{Basic data for the six stars\label{table-1}\vspace{2mm}}
\begin{tabular}{@{}crcrcrllcc@{}}
\hline
\\[-8pt]
CABS3 & HD$/$BD &  VS desig. & V~~ & (B\,$-$\,V) & Parallax~~~ & ~~$M_{V}$ & Sp. type & N &\\
      &         &       & m~~ &  m  &  arc ms~~~~  &  ~~ m     &          &   &\\[5pt]
\hline\hline \\[-5pt]
~~161&82159&GS Leo&8.85~&0.92&21.11\hspace{2mm}$\pm$\hspace{1.5mm}5.71&+5.5&~~G9\hspace{1mm}V&30\\
~~177&89959&---&8.30~&0.68&---\hspace{7mm}~&\hskip0.6em---&~~K0\hspace{1mm}V&33\\
~~216&\llap{+}$39^{\circ}$2587&---&9.27~&0.84&26.24\hspace{2mm}$\pm$\hspace{1.5mm}1.75&+6.4&~~G6\hspace{1mm}V\rlap{(?)}&82\\
~~257&138157&OX Ser&7.14&1.02&5.07\hspace{2mm}$\pm$\hspace{1.5mm}1.00&+0.7&~~K0\hspace{1mm}III&24\\
~~267&143705&---&7.96~&0.59&16.16\hspace{2mm}$\pm$\hspace{1.5mm}0.98&+4.0&~~G0\hspace{1mm}V&17\\
~~---\hspace{2mm}&160934&---&\llap{1}0.28&1.3:~~&40.75\hspace{2mm}$\pm$\hskip-0.05em 12.06&+8.3\rlap{:}&~~K8\hspace{1mm}V&38\\[5pt]
\tableline
\end{tabular}
\end{table*}

The new radial-velocity observations presented here were made with the 36-inch (0.91-m) coud\'{e} reflector at the Cambridge Observatories, England, with a photoelectric radial-velocity spectrometer operating on the cross-correlation principle that was first developed \citep{griffin67} at the same telescope by one of the present authors; the instrument in current use is largely patterned after the \textit{`Coravel'} designed by \citet{bmp79}. The radial-velocity traces (Fig.\,\ref{figure-2} below is an example) are cross-correlation functions that exhibit minima at abscissae corresponding to the velocities of the stars observed; in the case of a double-lined star there can be two minima (informally characterized here as `dips'). The dips may be thought of as averaged profiles of the absorption lines in the stellar spectrum.

The traces are routinely reduced by matching to digi\-tally synthesized models that incorporate rotational broadening, which is computed by starting with an empirical profile whose ordinates are scaled directly from observed traces which have the minimum observed half-width. Many stars share a very sharply defined lower bound to the dip width, which is taken as representing the zero-rotation profile. The spectral lines of active stars, such as those treated in this paper, often show considerable rotational broadening. It is quantified in the model dips by the summation of many elements, into which the stellar disc is conceptually divided, each of which is assigned the appropriate velocity and a brightness which is assessed according to a conventional limb-darkening law. The $v$\,sin\,$i$ values (projected rotational velocities) thus determined often repeat very well (to better than $\pm$~1~km\,s$^{-1}$ r.m.s.) from one observation to another of the same star, so the mean can easily be precise to a very few tenths, but in view of the neglect of non-rotational sources of broadening the external (true) uncertainty of the mean $v$\,sin\,$i$ determined in this fashion is never claimed to be better than 1 km\,s$^{-1}$.

Inasmuch as the stars treated here are all objects that exhibit chromospheric activity, they may (and at least in some cases do) have starspots.  The rotation of a spotted surface must affect, at some level, the radial velocities measured from the integrated light of the visible hemisphere of a star.  The effect is not large enough in the objects treated here, however, to vitiate the radial-velocity curves significantly, or even to add appreciably to the uncertainties of their determination.

\section{Orbits}
\subsection{HD~82159 (GS Leo)}

\begin{figure}[h]
\begin{center}
\includegraphics[scale=0.35, angle=-90]{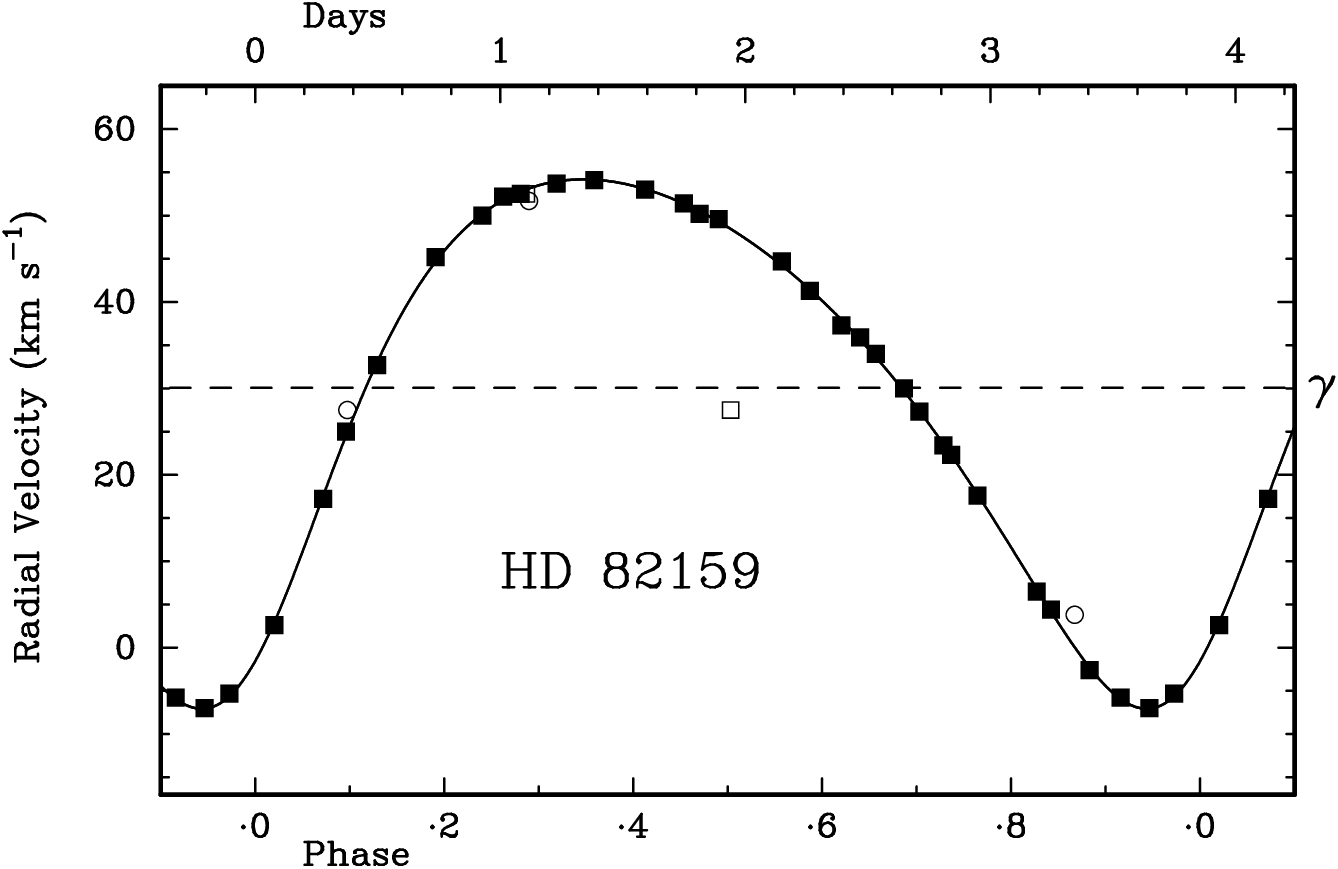}
\caption {The observed radial velocities of HD~82159 plotted as a function of phase, with the velocity curve
corresponding to the adopted orbital elements drawn through them. The observations recently made with
the Cambridge \textit{Coravel} are plotted as filled squares; the three by \citet{cutis99} are shown as open
circles and the two by \citet{strass00} as open squares. The `wild' one has of course been rejected
from the solution of the orbit; it is almost certainly an observation of the visual companion, ADS 7406 B.\label{figure-1}}
\end{center}
\end{figure}

This star, called `BD\,+11\ 2052\,A' by GMFL, is the primary in a wide (14\arcsec) visual binary, ADS 7406 \citep{aitken32}. The system is certainly a physical double star: the common proper motion of its components, long obvious, has been specifically pointed out by \citet{vander} and \citet{halb}. The components are of very similar brightness---indeed, to the eye sometimes one and sometimes the other appears to be the brighter, and a glance at the \textit{Hipparcos} `epoch photometry' demonstrates that that may very well be objectively so. The 30 new Cambridge observations, all made on separate nights in the 2008/9 observing season, are listed in Table~\ref{table-2}, at the head of which are included the three radial velocities published by \citet{cutis99} and two given by \citet{strass00}. An orbit computed from the Cambridge observations alone has a period of 3.85609~$\pm$~0.00015 days. During the 1300-odd cycles since the \citeauthor{cutis99} epoch, the uncertainty in the period produces a phasing uncertainty of about 0.2 days---still small in comparison with the orbital period, so we can be certain of the cycle count. Inclusion of the published velocities (those of \citeauthor{cutis99} with the very small weight of 0.02 so as to bring their variance into near-equality with that of Cambridge data; the \citeauthor{strass00} ones given an adjustment of +0.8 km\,s$^{-1}$ and a weighting of {\small$^{1}\hskip-0.3em/_{\hskip-0.2em4}$}, as was found appropriate to them by \citet{griffin09}) then produces the orbit whose elements are given in Table~\ref{table-3} and which is plotted in Fig.\,\ref{figure-1}. The period was changed from the Cambridge value by 2 units in the fourth decimal place---a little over one standard deviation. When the orbit was being solved, it was found necessary to reject one of the two \citeauthor{strass00} observations, which was far off the velocity curve. Its velocity makes it practically certain that it really belongs to the other component of the visual binary---its authors must have observed the wrong star.

\begin{table}[h!]
\setlength{\doublerulesep}{\arrayrulewidth}
\footnotesize 
\setlength\tabcolsep{1mm}
\caption{Radial-velocity observations of HD~82159 \label{table-2}}
{Except as noted, the observations were made at Cambridge}\\[5pt]
\begin{tabular}{@{}rrrrr@{}}
\tableline 
\\[-5pt]\hspace{1.5mm}
Date (UT)~~~~~~~~&        MJD\footnotemark~~\hspace{3.8em}&   Velocity~&   Phase\footnotemark\,\hspace{1.4em}&    (O -- C)~\,\\
           &         &     km\  s$^{-1}$~&          &   km\  s$^{-1}$~\,\\[5pt]
\hline\hline\\[-5pt]
1995 Jan.\hspace{0.5em}22.214$^{*}$&  49739.214 &     +51.7~~ &   $\overline{1319}.290$ & --1.4~~~~~\\
     25.329$^{*}$&    742.329 &     +27.5~~ &   $\overline{1318}.098$ &  +2.6~~~~~\\
     28.297$^{*}$&    745.297 &      +3.8~~ &       .867 &      +3.8~~~~~\\
           &            &            &            &            \\
1999 Feb.\hspace{0.5em}23.263$^{\ddagger}$&  51232.263&  +27.5~~ &  $\overline{932}.503$ & --21.0~~~~~\\
     26.286$^{\dagger}$&    235.290 &     +52.5~~ &    $\overline{931}.287$ &  --0.5~~~~~\\
           &            &            &            &            \\
2008 Dec.\hspace{0.5em}27.194~ &  54827.190 &      +6.5~~ &      0.827 &     --0.3~~~~~\\
           &            &            &            &            \\
2009 Jan.\hspace{1em}3.109~ &  54834.110 &     +37.3~~ &      2.621 &     --0.6~~~~~\\
      6.163~ &    837.160 &     +53.0~~ &      3.413 &     --0.1~~~~~\\
     14.097~ &    845.100 &     +50.2~~ &      5.470 &     --0.3~~~~~\\
     21.079~ &    852.080 &     +52.5~~ &      7.281 &     --0.3~~~~~\\
     24.129~ &    855.130 &     +17.2~~ &      8.072 &     --0.4~~~~~\\
Feb.\hspace{1em}4.095~ &   866.100 &     --5.8~~ &     10.916 &      +0.1~~~~~\\
     7.130~ &    869.130 &     +27.3~~ &     11.703 &     --0.1~~~~~\\
     8.068~ &    870.070 &      --7.0~~ &       .946 &      +0.1~~~~~\\
     11.083~ &    873.080 &     +23.4~~ &     12.728 &     --0.2~~~~~\\
     12.025~ &    874.030 &     --5.3~~ &       .973 &      +0.3~~~~~\\
     14.022~ &    876.020 &     +49.6~~ &     13.490 &      +0.3~~~~~\\
     17.000~ &    879.000 &     +52.2~~ &     14.262 &      +0.4~~~~~\\
Mar.\hspace{1em}6.058~ &  896.060 &     +30.0~~ &     18.687 &      +0.3~~~~~\\
     20.983~ &    910.980 &     +44.7~~ &     22.557 &      +0.5~~~~~\\
     26.916~ &    916.920 &     +25.0~~ &     24.096 &      +0.5~~~~~\\
     27.929~ &    917.930 &     +54.1~~ &       .359 &       0.0~~~~~\\
     29.014~ &    919.010 &     +35.9~~ &       .640 &      +0.2~~~~~\\
     29.951~ &    919.950 &     --2.6~~ &       .883 &     --0.3~~~~~\\
Apr.\hspace{1em}1.934~ &  922.930 &     +34.0~~ &     25.657 &      +0.4~~~~~\\
      7.849~ &    928.850 &     +45.2~~ &     27.191 &      +0.5~~~~~\\
      8.862~ &    929.860 &     +51.4~~ &       .454 &       0.0~~~~~\\
      9.953~ &    930.950 &     +22.3~~ &       .736 &       0.0~~~~~\\
     19.911~ &    940.910 &     +53.7~~ &     30.319 &     --0.2~~~~~\\
     20.944~ &    941.940 &     +41.3~~ &       .587 &     --0.2~~~~~\\
     21.928~ &    942.930 &      +4.4~~ &       .842 &      +0.2~~~~~\\
May\hspace{1em}8.887~ &  959.890 &     +50.0~~ &     35.240 &     --0.2~~~~~\\
     10.908~ &    961.910 &     +17.6~~ &       .764 &     --0.1~~~~~\\
     11.894~ &    962.890 &      +2.6~~ &     36.020 &     --0.3~~~~~\\
     23.881~ &    974.880 &     +32.7~~ &     39.129 &     --0.3~~~~~\\[3pt]
\tableline
\end{tabular}
\\[5pt] $^{*}$Observed by \citet{cutis99}; weight 0.02.
\\ $^{\dagger}$Observed by \citet{strass00}; weight {\footnotesize $^{1\hskip-0.3em}/_{\hskip-0.2em4}$}.
\\ $^{\ddagger}${Observed by \citet{strass00}; rejected.}
\end{table}

As a matter of interest, that other component was (deliberately) observed at Cambridge on six nights, with the results given in Table~\ref{table-4}. It shows no evidence of variation over the few months covered by the observations; its mean velocity is +28.75~$\pm$~0.12 km\,s$^{-1}$. There appears, therefore, to be a difference of 1.32~$\pm$~0.14 km\,s$^{-1}$ between the components of the visual binary. Their parallaxes were singularly poorly determined by \textit{Hipparcos}, but in round numbers the stars are 40 pc away, so their {14\arcsec} apparent separation represents a projected distance of about 560 AU. Orbital velocities scale as the square root of the orbital radii in circular orbits, so if the Earth were removed to 560 AU from the Sun its orbital velocity of 30 km\,s$^{-1}$ would be reduced to 30/$\sqrt{560}$ or $\sim$1.26 km\,s$^{-1}$. Therefore it is quite admissible for the component stars to differ in radial velocity by the observed amount, especially since the ADS~7406 system doubtless contains more than one solar mass.

\footnotetext[1]{MJD stands for Modified Julian Date, equal to JD $-$ 2,400,000.5 \citep{iau}.}
\footnotetext[2]{The phases include, as recommended in \citet{mcallister}, an integer part that represents a cycle count; negative cycle counts are indicated by overbars over the integer parts of the phases concerned.}

\begin{table}[h!]\begin{center}
\setlength{\doublerulesep}{\arrayrulewidth}
\caption{Orbital elements for HD~82159 \label{table-3}\vspace{2mm}}
\begin{tabular}{@{}ll@{}}
\hline\hline \\[-5pt]
\hskip1.5em$P$      &=~ 3.855881~$\pm$~0.000021 days$^{*}$\hskip 1.5 true cm \\
\hskip1.5em$T_{}$ &=~ MJD 54862.563~$\pm$~0.007 \\
\hskip1.5em$\gamma$ &=~ +30.07~$\pm$~0.06 km\,s$^{-1}$ \\
\hskip1.5em$K$  &=~ 30.63~$\pm$~0.10 km\,s$^{-1}$  \\
\hskip1.5em$e$      &=~ 0.2593~$\pm$~0.0029             \\
\hskip1.5em$\omega$ &=~ 214.9~$\pm$~0.7 degrees         \\
\hskip1.5em$a_{1}$\,sin\,$i$ &=~ 1.568~$\pm$~0.005 Gm \\ 
\hskip1.5em$f(m)$ &=~ 0.01036~$\pm$~0.00010 $M_{\odot}$ \\[5pt]
\multicolumn{2}{l}{~R.m.s. residual (wt.~1)~=~0.31 km\,s$^{-1}$} \\[3pt]
\tableline
\end{tabular}\\[5pt]\end{center}
{\small $^{*}$The true period, in the rest-frame of the system, is 3.855495~$\pm$~0.000021 days. It differs from the observed period because the velocity of recession of the system (the $\gamma$-velocity) lengthens the wavelengths of its radiation by the factor $\gamma$/c, which in this case amounts to 18 standard deviations (of the true period).  The significance of the true period is that it is the one from which the `derived elements' $a$\,sin\,$i$ and $f(m)$ should be calculated.}
\end{table}
 
The second visual component was also observed by \citet{cutis99}, whose three mutually concordant measurements have a mean of +26.8~$\pm$~0.3 km\,s$^{-1}$; the discrepancy from the Cambridge value may be evidence of a difference of zero-point, but it is of the opposite sign to the discrepancy implicit in the residuals of the \citeauthor{cutis99} observations of the visual primary. Two radial velocities of each component are reported by \citet{nordstrom}, but they are not given individually. (The components are misidentified in that listing, their designations being interchanged.) The mean for the constant-velocity star is given as +27.3~$\pm$~0.2 km\,s$^{-1}$.

\begin{table}[h!]
\setlength{\doublerulesep}{\arrayrulewidth}
\small
\center
\caption{Cambridge radial velocities of ADS 7406 B \label{table-4}\vspace{2mm}}
\begin{tabular}{@{}rcl@{}}
\hline \\[-5pt]
UT Date~~~~ & Velocity&\\
        &  ~km s$^{-1}$\\[5pt]
\hline\hline \\[-5pt] 
~2008 Dec.~~27.19 & +29.0 \\
2009 Jan.~~ 3.11 & +29.0 \\
Feb.~~ 7.13 & +28.8 \\
11.08 & +28.2 \\
17.00 & +28.7 \\
Mar.~~27.93 & +28.8 \\[3pt]
\hline
\end{tabular}
\end{table}

GMFL gave for HD~82159 a period of 3.8562 days---extremely close to ours, although they evidently did not avail themselves of the published radial velocities, and it is a mystery how they could have obtained such precision in a single observing run lasting only twelve days. Apart from the period, the only other information they gave about the orbit is that it is `eccentric'. That is, in fact, quite unusual in a system having such a short period. The mean $v$\,sin\,$i$ value is 14.2~$\pm$~0.15 km\,s$^{-1}$; it is given as 13~$\pm$~2 km\,s$^{-1}$ by \citet{cutis99} and as 13 km\,s$^{-1}$ by \citet{nordstrom}. It may well be assumed that the rotation is pseudo-synchronized \citep{hut} to the orbital revolution, in which case the rotation period is shorter than the orbital one by a factor of about 1.42, making it 2.64 days. The projected rotational velocity then corresponds to a projected radius of 0.74 $R_{\odot}$. That is smaller by about 10--15\% than the actual radius that stars of the relevant type (G8\,V) are supposed to possess, so we might expect the inclination to be something like $60\arcdeg$. The uncertainty of the real radius, together with the great sensitivity of $i$ to sin\,$i$ at high inclinations, conspire to make the estimation of $i$ very uncertain. The projected rotational velocity of the B component is found from the Cambridge traces to be 3.9~$\pm$~0.4 km\,s$^{-1}$ (\citeauthor{cutis99} 6 km\,s$^{-1}$, \citeauthor{nordstrom} (with stellar identities corrected) 4\ km\,s$^{-1}$).

The mass function includes a sin$^{3}\,i$ term, which according to the above estimate of $i$ may be expected to be about 0.65, leaving us with $m_{2}^{3}\,/\,(m_{1}\,+\,m_{2})^{2}\sim 0.016\ M_{\odot}$. With a conventional mass of something like $0.9\ M_{\odot}$ for $m_{1}$, the mass of the unseen secondary, $m_{2}$, is found to be just under 0.3 $M_{\odot}$---well down into the M-dwarf sequence. GMFL suggested a secondary of type K5\,V, but it seemed that they could model the spectrum just as well without it, and it now appears preferable to suppose the secondary to be a good deal cooler and fainter than K5.

\subsection{HD~89959}

\begin{figure}[h]
\begin{center}
\includegraphics[scale=0.35, angle=-90]{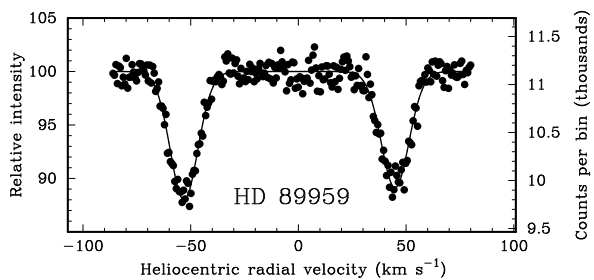}
\caption[] {Radial-velocity trace of HD~89959, obtained with the Cambridge \textit{Coravel} on 2009 May 17, illustrating
the widely separated double lines.\label{figure-2}}
\end{center}
\end{figure}

\begin{figure}[h!]
\begin{center}
\includegraphics[scale=0.35, angle=-90]{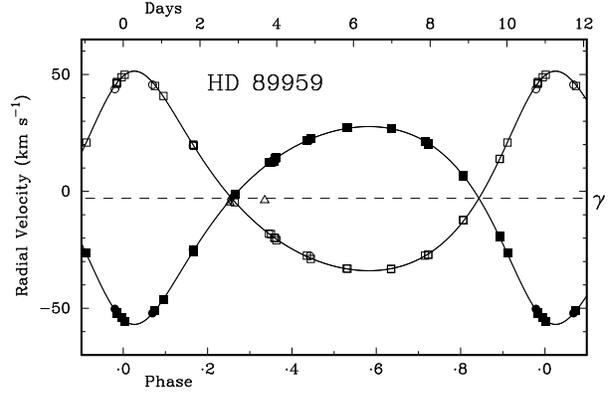}
\caption[] {The observed radial velocities of HD~89959 plotted as a function of phase, with the velocity curves corresponding to the adopted orbital elements drawn through them. Cambridge observations are plotted as squares, those of \citet{wsh03} as circles; the primary and secondary components of the binary are distinguished by filled and open symbols respectively.  Observations reduced as single-lined when the components were blended together are represented by open triangles (Wichmann et al.) and open diamonds (Cambridge; more or less hidden near phase .25). The bunching of the points arises from the proximity of the period to an integral number of days.\label{figure-3}}
\end{center}
\end{figure}

HD\,89959 is a double-lined object with components that are very similar to one another. An example of a radial-velocity trace of it, obtained when the dips were well separated, appears as Fig.\,\ref{figure-2}. There are 33 recent Cambridge observations, listed in Table~\ref{table-5} together with six measurements obtained at the KPNO coud\'{e} feed and published by \citet{wsh03}. Observations obtained when the spectra of the components were blended together are listed between the columns for the primary and the secondary and were not taken into account in the solution of the orbit. The Cambridge data alone yield an orbital period of 10.9927~$\pm$~0.0005 days. At the epoch of the Wichmann et al. observations, some 300 cycles earlier, the uncertainty in the phasing has grown to about 0.17 days---still very small in comparison with the orbital period---so there is absolutely no doubt as to how many cycles have elapsed since that epoch. Accordingly, the \citeauthor{wsh03} measures can be included in the solution with full confidence to refine the period; with a view to bringing them into approximate homogeneity with the Cambridge velocities and to equalize the variances of the two series they have been given an empirical adjustment of +1.5 km\,s$^{-1}$ and attributed a weight of {\small$^{1}\hskip-0.3em/_{\hskip-0.2em4}$}. The resulting orbital elements are shown in Table~\ref{table-6}, and the orbit is plotted in Fig.\,\ref{figure-3}. Owing to the nearness of the period to the integral number of 11 days, the data points appear in 11 bunches; that has an interesting cosmetic effect but does not influence the accuracy or merit of the derived orbit. There are multiple observations at all of the 11 phases that were accessible during the observing campaign. By chance, the phases of the \citeauthor{wsh03} measures, which were all made in one observing run, are close to those of the Cambridge ones.

\begin{table*}\clearpage
\setlength{\doublerulesep}{\arrayrulewidth}
{\footnotesize
\setlength\tabcolsep{1mm}
\caption{Radial-velocity observations of HD~89959 \label{table-5}}
{Except as noted, the observations were made at Cambridge}\\[5pt]
\begin{tabular}{@{}rrrrrrr@{}}
\tableline\\[-5pt]
Date (UT)~~~~~~& MJD~~ & \multicolumn{2}{c}{Velocity}~ & Phase~~&\multicolumn{2}{c}{(O -- C)} \\
       &      & Prim.~ &  Sec.~~~&       &  Prim.~~&  Sec.~~~\\
     &      & km\,s$^{-1}$ &  km\,s$^{-1}$&       &  \llap{k}m\,s$^{-1}$ &  km\,s$^{-1}$ \\[5pt]

\hline\hline\\[-5pt]
 1999\hspace{0.5em}Mar.\hspace{0.5em}27.25$^{*}$&51264.25& --\,50.3~~&+43.8~~&~~$\overline{325}$.979&    +0.4~~~&    --\,1.4~~~~\\
28.23$^{*}$&    265.23&    --\,52.1~~&    +45.6~~&~~$\overline{324}$.069&    +0.1~~~&   --\,1.1~~~~\\
     29.29$^{*}$&    266.29&   --\,24.9~~&    +19.7~~&.165&    +0.8~~~&    --\,0.3~~~~\\
     30.25$^{*}$&    267.25&  \multicolumn{2}{c}{--4.8}&.252&---~~~~&---~~~~~\\
     31.16$^{*}$&    268.16&  \multicolumn{2}{c}{--3.8}&.335&---~~~~&---~~~~~\\
Apr.\hspace{1em}1.34$^{*}$&    269.34&    +22.0~~&    --\,27.6~~&.442&    --\,0.3~~~&    +0.8~~~~\\
&&&&&&~~~~\\
 2008\hspace{0.5em}Dec.\hspace{0.5em}27.22~\,&  54827.22&    --\,46.1&    +40.8~~&     0.095&    --\,0.1~~~&    +0.3~~~~\\
&&&&&&~~~~\\
 2009\hspace{0.5em}Jan.\hspace{1em}3.13~\,& 54834.13&    +20.2~~&    --\,27.1~~&     0.723&    --\,0.4~~~&    --\,0.3~~~~\\
      6.20~\,&   837.20&    --\,55.6~~&    +49.9~~&     1.003&    --\,0.3~~~&    +0.1~~~~\\
     14.13~\,&   845.13&    +20.4~~&    --\,27.1~~&      .724&    --\,0.2~~~&    --\,0.4~~~~\\
     19.95~\,&   850.95&      \multicolumn{2}{c}{--3.5}&     2.253&---~~~~&---~~~~~\\
     21.12~\,&   852.12&    +14.0~~&    --\,20.0~~&.360&    +0.1~~~&     0.0~~~~\\
     24.15~\,&   855.15&    +27.0~~&    --\,33.1~~&.635&    +0.2~~~&    --\,0.1~~~~\\
Feb.\hspace{1em}4.14~\,&  866.14&    +27.0~~&    --\,33.3~~&     3.635&    +0.2~~~&    --\,0.2~~~~\\
      7.18~\,&   869.18&    --\,26.5~~&    +20.9~~&.912&    +0.3~~~&    --\,0.2~~~~\\
      8.10~\,&   870.10&    --\,54.0~~&    +48.8~~&.995&    +0.2~~~&    +0.1~~~~\\
     12.14~\,&   874.14&    +14.5~~&    --\,20.9~~&     4.363&    +0.2~~~&    --\,0.5~~~~\\
     17.01~\,&   879.01&     +7.0~~&    --\,12.2~~&.806&    +0.1~~~&    +0.8~~~~\\
Mar.\hspace{0.8em}5.04~\,&   895.04&     --\,1.4~~&     --\,4.7~~&     6.264&    +0.1~~~&    --\,0.2~~~~\\
      6.07~\,&   896.07&    +13.0~~&    --\,19.8~~&.358&    --\,0.7~~~&     0.0~~~~\\
     21.00~\,&   911.00&    +21.2~~&    --\,27.5~~&     7.716&    --\,0.2~~~&    +0.1~~~~\\
     23.95~\,&   913.95&    --\,51.9~~&    +46.5~~&.984&     0.0~~~&    +0.1~~~~\\
     26.95~\,&   916.95&    \multicolumn{2}{c}{--2.9}&     8.257&---~~~~&---~~~~~\\
     27.97~\,&   917.97&    +12.4~~&    --\,18.3~~&.350&    --\,0.3~~~&    +0.5~~~~\\
     29.01~\,&   919.01&    +22.4~~&    --\,28.9~~&.445&     0.0~~~&    --\,0.3~~~~\\
     29.97~\,&   919.97&    +27.1~~&    --\,33.1~~&.532&    +0.1~~~&    +0.1~~~~\\
Apr.\hspace{1em}2.00~\,&   923.00&     +6.4~~&    --\,12.4~~&.808&    --\,0.1~~~&    +0.2~~~~\\
      5.93~\,&   926.93&    --\,25.7~~&    +19.9~~&     9.165&     0.0~~~&     0.0~~~~\\
      7.91~\,&   928.91&    +12.2~~&    --\,18.1~~&.345&    +0.2~~~&     0.0~~~~\\
      8.90~\,&   929.90&    +21.9~~&    --\,27.5~~&.435&    +0.2~~~&    +0.4~~~~\\
     20.93~\,&   941.93&    +27.1~~&    --\,33.0~~&    10.530&    +0.2~~~&    +0.1~~~~\\
     24.93~\,&   945.93&    --\,19.6~~&    +13.9~~&.894&    +0.2~~~&    --\,0.1~~~~\\
May\hspace{1em}6.92~\,&   957.92&    --\,52.1~~&    +46.8~~&    11.984&    --\,0.2~~~&    +0.4~~~~\\
      8.93~\,&   959.93&    --\,25.1~~&    +19.5~~&    12.167&     0.0~~~&    +0.1~~~~\\
     16.91~\,&   967.91&    --\,19.3~~&    +13.9~~&.893&    +0.3~~~&    +0.1~~~~\\
     17.90~\,&   968.90&    --\,51.9~~&    +46.1~~&.983&    --\,0.3~~~&     0.0~~~~\\
     28.90~\,&   979.90&    --\,52.0~~&    +46.1~~&    13.984&    --\,0.2~~~&    --\,0.2~~~~\\
     29.90~\,&   980.90&    --\,50.8~~&    +45.1~~&    14.075&    +0.2~~~&    --\,0.3~~~~\\
     30.90~\,&   981.90&    --\,25.7~~&    +20.0~~&.166&    --\,0.2~~~&    +0.2~~~~\\
\tableline
\end{tabular}
\\[5pt] $^{*}$Observed by \citet{wsh03}; weight {\scriptsize$^{1}\hskip-0.3em/_{\hskip-0.15em4}$}.}
\end{table*}
There is an outright conflict between the period of 10.9929 days given here and that of 12.1606 days asserted by GMFL. Although they did not publish their observations, GMFL reported that there were three of them, all made in one observing run, whose dates they gave, that lasted barely one cycle of HD~89959's orbit. The GFML orbit was based just on \citeauthor{wsh03}'s six measures (two of which were blends) and their own. The former set, made on six consecutive nights, did not permit its authors to hazard even a preliminary estimation of the orbit, and GMFL certainly could not have obtained an orbit from their own three points---so there seems to be no means whereby they could have decided on the number of cycles that intervened between the Wichmann observations and their own. The number that they selected was evidently 151, whereas the actual number of intervening cycles was 167, and their period is in error by that proportion. 

\begin{table}[h!]
\setlength{\doublerulesep}{\arrayrulewidth}
\begin{center}
\caption{Orbital elements of HD~89959 \label{table-6}\vspace{2mm}}
\begin{tabular}{@{}ll@{}}
\hline\hline\\[-5pt]
\hskip1.5em$~P$      &=~ 10.99291~$\pm$~0.00003 days$^{*}$ \\
\hskip1.5em$~T_{}$ &=~ MJD 54815.186~$\pm$~0.007 \\
\hskip1.5em$~\gamma$ &=~ --2.97~$\pm$~0.03 km\,s$^{-1}$ \\
\hskip1.5em$~K_{1}$  &=~ 42.31~$\pm$~0.07 km\,s$^{-1}$  \\
\hskip1.5em$~K_{2}$  &=~ 42.68~$\pm$~0.07 km\,s$^{-1}$  \\
\hskip1.5em$~q$      &=~ 1.0088~$\pm$~0.0023 (=$m_{1}/m_{2}$) \\
\hskip1.5em$~e$      &=~ 0.2877~$\pm$~0.0011             \\
\hskip1.5em$~\omega$ &=~ 162.60~$\pm$~0.24 degrees         \\
\hskip1.5em$~a_{1}$\,sin\,$i$ &=~ 6.125~$\pm$~0.010 Gm \\ 
\hskip1.5em$~a_{2}$\,sin\,$i$ &=~ 6.179~$\pm$~0.010 Gm \\ 
\hskip1.5em$~f(m_{1})$ &=~ 0.0760~$\pm$~0.0004 $M_{\odot}$ \\
\hskip1.5em$~f(m_{2})$ &=~ 0.0780~$\pm$~0.0004 $M_{\odot}$ \\
\hskip1.5em$~m_{1}$\,sin$^{3}\,i$ &=~ 0.3092~$\pm$~0.0013 $M_{\odot}$ \\
\hskip1.5em$~m_{2}$\,sin$^{3}\,i$ &=~ 0.3065~$\pm$~0.0012 $M_{\odot}$ \\[5pt]
\multicolumn{2}{l}{~R.m.s. residual (unit weight) = 0.26 km\,s$^{-1}$} \\[3pt]
\hline
\end{tabular}\\[5pt]\end{center}
$^{*}${\small The true period, in the rest-frame of the system, is 10.99302~$\pm$~0.00003 days. It differs from the observed period  by 3.3 standard deviations.}
\end{table}

It is seen from the orbital elements that there is a difference of just 0.88~$\pm$~0.23 per cent in the masses of the two stars. That is equivalent to a difference in spectral type of only about one-third of a sub-type; GMFL found the spectra of the components to be ``nearly identical'' and of type K0\,V, though at (B\,--\,V) = $0\textsuperscript{m}.68$ they seem remarkably blue for such a type. The minimum masses, $m$\,sin$^{3}\,i$, required by the orbit are about one-third of the actual masses to be expected of G5–--K0 main-sequence stars, so sin\,$i$ must be about $\sqrt[3]{\small^{1}\hskip-0.3em/_{\hskip-0.2em3}}$, which is nearly 0.7, suggesting $i$ $\sim$ $44\arcdeg$. The areas of the dip signatures in radial-velocity traces can be quantified in exactly the same way as equivalent widths of absorption lines in tracings of stellar spectra, except that the abscissae of the tracings---and therefore the units of the equivalent widths---are km\,s$^{-1}$ instead of \AA. In the case of HD~89959 the mean areas found for the primary and secondary dips are 1.99~$\pm$~0.02 and 1.91~$\pm$~0.02 km\,s$^{-1}$, respectively, so the difference is about 4\%. It is to be interpreted as a small difference in luminosity, of about $0\textsuperscript{m}.04$; it is in the same sense as the difference in mass, but implies an even smaller fraction of a spectral sub-type between the types of the two stars. The projected rotational velocities of the stars are too small to be accurately determinable; upper limits of about 2 km\,s$^{-1}$ can be placed upon them.

\subsection{BD +$39\arcdeg$\,2587}

This star is the fainter component of a wide visual binary, whose primary is HD\,112733. For simplicity we will call the components A and B, the one of principal interest here being B. They definitely constitute a physical binary system, since they share a large common proper motion that has retained them in the same relative positions since they were first catalogued and measured by Struve (1827 and 1837 respectively). The angular separation is about {35\arcsec}, with B in position angle $83\arcdeg$, $i.e.$\,almost due east of A. The separation is so wide that Aitken did not deign to include the system as a double star in his great catalogue \citep{aitken32} of such objects, although it did feature in the earlier one by \citet{burn06} as BDS 6331. Both stars were on the \textit{Hipparcos} programme, with nos. 63317 and 63322 for A and B respectively; the parallaxes are 22.50\,$\pm$\,1.45 and 26.24\,$\pm$\,1.75 arc-milliseconds---the difference is a bit surprising, but at 1.65 times the joint standard deviation it is marginally acceptable. The projected linear separation of the stars, at the implied distance of about 40 pc, is about 1400 AU. There is a difference of six tenths of a magnitude in V between them, their magnitudes being $8\textsuperscript{m}.67$ and $9\textsuperscript{m}.28$. The (B$\,-\,$V) colour indices are $0\textsuperscript{m}.74$ and $0\textsuperscript{m}.85$, respectively, the difference being quite noticeable when the system in viewed through a telescope. That makes it the more disconcerting that the spectra have been classified as K0\,V \citep{lopez} and G6\,V \citep{favata98}, respectively, making it appear that the redder star has considerably the earlier type. The distance modulus corresponding to the mean of the parallaxes is $3\textsuperscript{m}.07$, making the absolute V magnitudes $5\textsuperscript{m}.60$ and $6\textsuperscript{m}.21$. Such absolute magnitudes would normally \citep{allen73} characterize main-sequence types of about G8 and K1, respectively.

Active-chromosphere stars demonstrate their activity in various ways, including emission in the H and K lines of Ca II and often in H$\alpha$, and in the ultraviolet, including X-rays. The Einstein `Medium-Sensitivity' X-ray survey \citep{gioia90,stocke} detected a source, dubbed 1E1256.2+3833, about 2 seconds of time east of the position of B; it is to be recalled that A is about {35\arcsec}, or 3 seconds of time, to the west of B. The 90\%-confidence error circle of the X-ray position is {48\arcsec} in radius and includes the position of B but not of A. It is confusing that \citet{stocke}, in their list of optical identifications of the X-ray sources, say that the relevant source is a star of type K3\,V; they do not identify the star properly, but have a specific note saying ``This stellar ID is NOT SAO 63275, which is approximately 20\H{} west of the proposed ID and outside the X-ray error circle.'' In actuality the B component, which we feel sure is the active star, \textit{is} SAO 63275; it is {20\arcsec} west of the X-ray position and thus well inside the error circle; the only other star of comparable brightness anywhere near is A, which is {35\arcsec} west of what $we$ think must be the true ID, {55\arcsec} west of the X-ray position and therefore outside the error circle, but (like all stars other than B) fits the description of not being SAO 63275. Perhaps on account of that confusion, Simbad misidentifies 1E1256.2+3833 with HD\,112733.

\citet{favata95} determined rotational velocities and other properties for a number of stars and included 1E1256.2+3833 in a list in which it is specifically noted as belonging to a `normal' sample, from which chromospherically active stars were excluded, but nevertheless to have a $v$\,sin\,$i$ of 19 km\,s$^{-1}$---which would not characterize any but an active star, though in fact it is much greater than is exhibited by either the A or the B component of BDS\,6331. The \citeauthor{favata98} list actually has two identical entries for 1E1256.2+3833, on consecutive lines, but no reason for the duplication is apparent. We conclude that there has been considerable confusion in the literature between the components, as indeed seems to be the case also with the ADS\,7406 system treated above.

\begin{figure}[h]
\begin{center}
\includegraphics[scale=0.35, angle=-90]{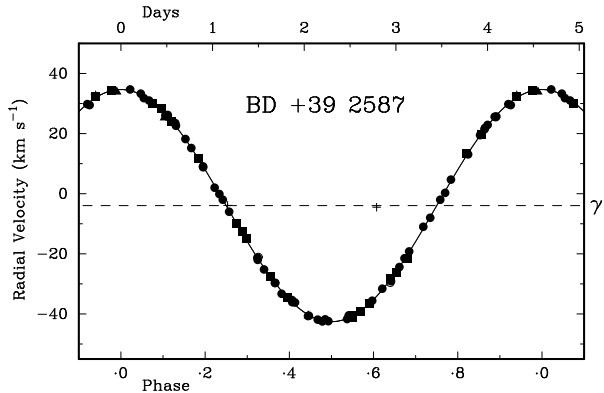}
\caption[] {The observed radial velocities of BD +39$\arcdeg$\,2587 plotted as a function of phase, with the velocity curve
corresponding to the adopted orbital elements drawn through them. The majority (56) of the observations,
plotted as filled circles, were made with the \textit{Coravel} at Haute-Provence, and there are two from its clone at ESO. There are 17 from the Cambridge \textit{Coravel} (filled squares), three from the DAO (filled triangles) and
one from Palomar (filled star). Two measurements by Strassmeier et al. (2000) are shown as plusses, but were
not included in the solution of the orbit because at least one of them is evidently of the visual companion
star, HD~112733. Symbols plotting three measurements made with the original radial-velocity spectro\-meter at Cambridge are completely hidden by other points.
\label{figure-4}}
\end{center}
\end{figure}

Unlike the other stars discussed in this paper, the components of BDS\,6331 have been on the Cambridge radial-velocity observing programme for a long time. Owing to its late spectral type and high Galactic latitude (about $79\arcdeg$), the A component was included in the comprehensive programme \citep{yoss} to measure radial velocities for all late-type Henry Draper Catalogue stars within $15\arcdeg$ of the North Galactic Pole. B was initially observed just out of interest as an obvious visual companion to A, and was then measured many times, mostly with the \textit{Coravel} at Haute-Provence \citep{bmp79}, after it proved to be a short-period spectroscopic binary. In the 30-odd years that the star has been under observation it has been measured a total of 82 times, not only at Cambridge and Haute-Provence but also with the instruments at Palomar \citep{griffin74}, the Dominion Astrophysical Observatory (DAO; \citeauthor{flet82} \citeyear{flet82}), and at ESO (a clone of the Haute-Provence \textit{Coravel}); and two measurements have been published by \citet{strass00}. The diversity of sources has required some alignment of zero-points and appropriate weighting, specified in footnotes to Table~\ref{table-7}, where the observations are set out. They lead to the orbital elements in Table~\ref{table-8}; the orbit is shown in Fig.\,\ref{figure-4}. The interest in the short-period binary seemed to warrant repeated measurements of its visual companion, HD\,112733, too, and those velocities are listed in Table~\ref{table-9}. Just as in the case of HD\,82159, it is found that one of the two observations by \citet{strass00} is far from the veloci\-ty curve but has the right veloci\-ty to be an observation of the other visual component, which it almost certainly is. The other observation from the same source has a similar velocity; it falls close to the velocity curve but could equally well be of the wrong star, so both have been rejected as velocities of B. 

\begin{table}\clearpage
\setlength\tabcolsep{1.6mm}
\setlength{\doublerulesep}{\arrayrulewidth}
{\footnotesize
\caption{Radial-velocity observations of BD+39$\arcdeg$\,2587 \label{table-7}}
{Made at OHP (offset +0.8 km\,s$^{-1}$, weight 1) except as noted}\\[5pt]
\begin{tabular}{@{}rrrrr@{}}
\hline \\[-5pt]
Date (UT)~~~~~ &        MJD~~~~ &   Velocity &      Phase~~&    (O -- C) \\
           &            &     km\,s$^{-1}$ &            &     km\,s$^{-1}$ \\[5pt]
\hline\hline\\[-5pt]
1977\hspace{0.5em}May\hspace{0.6em}28.94$^{*}$~~&  43291.94~\, &  +25.7~~&     0.110&   --\,0.1~~ \\
&&&&~~ \\
1986\hspace{0.5em}May\hspace{0.6em}28.98$^{*}$~~&  46578.98~\, &  --\,29.6~~&   715.640&   --\,1.0~~ \\
    \hspace{0.5em}Nov.\hspace{0.5em}25.52$^{\dagger}$~~&  759.52~\, &  +32.5~~&   754.940&   +0.5~~ \\
&&&&\\
1987\hspace{0.5em}Mar.\hskip1em4.134~~&  46858.134&  --\,35.4~~&   776.407&   +0.7~~ \\
&&&&\\
1988\hspace{0.5em}Mar.\hspace{0.5em}15.160~~&  47235.160&  --\,42.5~~&   858.478&   --\,0.3~~ \\
&&&&\\
1989\hspace{0.5em}Mar.\hspace{0.5em}25.100~~&  47610.100&  +28.3~~&   940.096&   +0.5~~ \\
     28.033~~&    613.033&   --\,8.0~~&.735&   --\,0.3~~ \\
     28.885~~&    613.885&  +29.8~~&.920&   --\,0.1~~ \\
     30.020~~&    615.020&  +15.2~~&   941.167&    0.0~~ \\
     30.149~~&    615.149&   +8.8~~&.195&   --\,0.3~~ \\
     30.932~~&    615.932&  --\,29.7~~&.366&   --\,0.1~~ \\
  Apr.\hskip1em8.94*~~&    624.94~\, &  --\,21.1~~&   943.327&   +0.7~~ \\
     27.866~~&    643.866&  --\,40.6~~&   947.446&   --\,0.2~~ \\
     28.078~~&    644.078&  --\,42.4~~&.493&   +0.1~~ \\
     28.960~~&    644.960&  --\,19.2~~&.685&   +0.2~~ \\
     29.904~~&    645.904&  +25.5~~&.890&   --\,0.3~~ \\
     30.068~~&    646.068&  +29.5~~&.926&   --\,1.0~~ \\
     30.894~~&    646.894&  +26.1~~&   948.106&   --\,0.4~~ \\
  May\hskip1em1.113~~&    647.113&  +18.2~~&.153&   +0.1~~ \\
      1.901~~&    647.901&  --\,22.0~~&.325&   --\,0.6~~ \\
      2.094~~&    648.094&  --\,29.7~~&.367&   +0.1~~ \\
      2.877~~&    648.877&  --\,41.7~~&.537&   --\,0.2~~ \\
&&&&\\
 1990\hspace{0.5em}Jan.\hspace{0.7em}27.047~~&  47918.047&  +22.6~~&  1007.131&   +0.3~~ \\
     31.141~~&    922.141&  +34.7~~&  1008.022&   +0.4~~ \\
  Feb.\hspace{0.5em}12.352$^{\ddagger}$&    934.352&  --\,21.7~~&  1010.680&   --\,1.3~~ \\
     14.369$^{\ddagger}$&    936.369&  +24.0~~&  1011.119&   --\,0.3~~ \\
  Mar.\hspace{0.5em}13.34$^{\S}$~~&    963.34~\, &  +33.9~~&  1016.990&   --\,0.7~~ \\
     15.43$^{\S}$~~&    965.43~\, &  --\,40.6~~&  1017.445&   --\,0.3~~ \\
&&&&~~ \\
 1991\hspace{0.5em}Jan.\hspace{0.7em}26.114~~&  48282.114&  --\,33.3~~&  1086.382&   --\,0.9~~ \\
     27.102~~&    283.102&  --\,35.6~~&.597&   +0.1~~ \\
     28.149~~&    284.149&  +13.1~~&.825&   --\,0.4~~ \\
     29.171~~&    285.171&  +33.3~~&  1087.047&   +0.3~~ \\
     30.066~~&    286.066&   --\,2.0~~&.242&    0.0~~ \\
     31.102~~&    287.102&  --\,41.9~~&.467&   --\,0.1~~ \\
  Feb.\hskip1.1em3.142~~&    290.142&  +23.5~~&  1088.129&   +0.9~~ \\
     4.110~~&    291.110&  --\,25.2~~&.340&   --\,0.6~~ \\
&&&&\\
 1992\hspace{0.5em}Jan.\hspace{0.7em}14.162~~&  48635.162&   --\,0.1~~&  1163.234&   --\,0.1~~ \\
     15.129~~&    636.129&  --\,40.7~~&.444&   --\,0.5~~ \\
     16.126~~&    637.126&  --\,24.4~~&.661&    0.0~~ \\
     17.165~~&    638.165&  +25.6~~&.888&   +0.2~~ \\
     18.193~~&    639.193&  +26.2~~&  1164.111&   +0.6~~ \\
     19.170~~&    640.170&  --\,21.3~~&.324&    0.0~~ \\
     20.167~~&    641.167&  --\,40.5~~&.541&   +0.8~~ \\
     21.219~~&    642.219&   +0.3~~&.770&   --\,0.6~~ \\
  Feb.\hspace{0.5em}28.50$^{\S}$~~&    680.50~\, &  +25.3~~&  1173.103&   --\,1.5~~ \\
  Apr.\hspace{0.5em}22.042~~&    734.042&   --\,2.0~~&  1184.758&    0.0~~ \\
     24.048~~&    736.048&   +9.1~~&  1185.195&    0.0~~ \\
     25.051~~&    737.051&  --\,36.2~~&.413&   +0.8~~ \\
     26.099~~&    738.099&  --\,29.1~~&.641&   --\,0.8~~ \\
     27.071~~&    739.071&  +19.5~~&.853&   +0.2~~ \\
     29.974~~&    741.974&  --\,41.8~~&  1186.485&   +0.6~~ \\
   May\hskip1em1.045~~&    743.045&  --\,11.0~~&.718&   +0.7~~ \\
\tableline
\end{tabular}}
\end{table}
\begin{table}
\setlength\tabcolsep{1.5mm}
\setlength{\doublerulesep}{\arrayrulewidth}
{\footnotesize
\textbf{Table 7} (concluded)\\[5pt]
\begin{tabular}{@{}rrrrr@{}}
\hline \\[-5pt]
Date (UT)~~~~~ &        MJD~~~~ &   Velocity &      Phase~~&    (O -- C) \\
           &            &     km\,s$^{-1}$ &            &     km\,s$^{-1}$ \\[5pt]
\hline\hline\\[-5pt]
 1993\hspace{0.5em}Feb.\hspace{0.5em}15.131~~~&  49033.131&    +21.9~~&  1249.865&   +0.4~~ \\
  Mar.\hspace{0.3em}20.160~~~&    066.160&    +31.8~~&  1257.054& --\,0.6~~ \\
  Dec.\hspace{0.5em}28.229~~~&    349.229& --\,21.5~~&  1318.674&    +0.3~~ \\
&&&&~~ \\
 1994\hspace{0.5em}Feb.\hspace{0.5em}21.113~~~&  49404.113& --\,31.6~~&  1330.621&      +0.4~~ \\
  Apr.\hspace{0.5em}30.038~~~&    472.038& --\,36.1~~&  1345.407&    0.0~~ \\
  May\hskip0.7em~3.070~~~&    475.070& +31.0~~&  1346.067&   --\,0.3~~ \\
  Dec.\hspace{0.5em}31.244~~~&    717.244&  +4.7~~&  1398.784&   +0.5~~ \\
&&&&~~ \\
 1995\hspace{0.5em}Jan.\hskip1em5.238~~~&  49722.238&  +22.9~~&  1399.871&   +0.3~~ \\
  June\hspace{.8em}3.015~~~&    871.015&   --\,6.0~~&  1432.257&   --\,0.3~~ \\
  Dec.\hspace{0.5em}27.189~~~&  50078.189&  --\,27.6~~&  1477.355&    0.0~~ \\
&&&&~~ \\
 1996\hspace{0.5em}Apr.\hspace{0.9em}1.051~~~&  50174.051&   +2.0~~&  1498.223&   --\,0.7~~ \\
&&&&~~ \\
 1997\hspace{0.5em}Apr.\hspace{0.5em}11.067$^{\P}$&  50549.067&  +19.7~~&  1579.857&   --\,0.4~~ \\
  May\hspace{1em}5.041$^{\P}$&    573.041&  +30.1~~&  1585.076&   --\,0.3~~ \\
  July\hspace{0.5em}25.885~~~&    654.885&  +25.7~~&  1602.892&   --\,0.3~~ \\
&&&&~~ \\
 1998\hspace{0.5em}July\hspace{1em}9.889~~~&  51003.889&  +21.4~~&  1678.863&   +0.1~~ \\
&&&&~~ \\
 1999\hspace{0.5em}Feb.\hspace{0.5em}23.405$^{\parallel}$&  51232.405&   --\,4.5~~&  1728.607&  +29.6~~ \\
     26.373$^{\parallel}$&    235.373&   --\,3.9~~&  1729.253&   +0.9~~ \\
&&&&~~ \\
 2000\hspace{0.5em}Apr.\hspace{0.5em}22.022$^{\P}$&  51656.022&  +13.3~~&  1820.821&   +0.6~~ \\
  June\hspace{0.4em}19.972$^{\P}$&    714.972&  --\,26.1~~&  1833.654&   --\,0.1~~ \\
&&&&~~ \\
 2001\hspace{0.5em}Feb.\hspace{0.6em}27.159$^{\P}$&  51967.159&  --\,41.2~~&  1888.550&   --\,0.5~~ \\
&&&&~~ \\
 2002\hspace{0.5em}Mar.\hspace{0.4em}27.967$^{\P}$&  52360.967&  --\,10.0~~&  1974.275&    0.0~~ \\
     28.069$^{\P}$&    361.069&  --\,14.9~~&.297&   +0.4~~ \\
  Apr.\hspace{0.5em}19.998$^{\P}$&    383.998&  --\,12.6~~&  1979.289&   +0.6~~ \\
&&&&~~ \\
 2003\hspace{0.5em}Mar.\hspace{0.4em}17.060$^{\P}$&  52715.060&  --\,27.4~~&  2051.355&   +0.2~~ \\
  May\hspace{1em}7.903$^{\P}$&    766.903&  --\,28.1~~&  2062.640&   +0.4~~ \\
&&&&~~ \\ 
 2004\hspace{0.5em}May\hspace{0.5em}22.969$^{\P}$&  53147.969&  --\,36.4~~&  2145.591&    0.0~~ \\
&&&&~~ \\ 
 2005\hspace{0.5em}Jan.\hspace{0.5em}23.219$^{\P}$&  53393.219&  +34.4~~&  2198.978&   +0.1~~ \\
&&&&~~ \\
 2006\hspace{0.5em}June\hspace{0.3em}28.947$^{\P}$&  53914.947&  --\,40.6~~&  2312.549&   +0.2~~ \\
&&&&~~ \\
 2007\hspace{0.5em}Mar.\hspace{0.4em}27.079$^{\P}$&  54186.079&  --\,39.3~~&  2371.569&  --\,0.3~~ \\
&&&&~~ \\
 2008\hspace{0.5em}May\hspace{0.5em}20.947$^{\P}$&  54606.947&  +11.8~~&  2463.185&   +0.4~~ \\
&&&&~~ \\
 2009\hspace{0.5em}Mar.\hspace{0.4em}21.111$^{\P}$&  54911.111&  --\,34.6~~&  2529.396&    0.0~~ \\
     \hspace{0.5em}May\hspace{0.6em}26.921$^{\P}$&    977.921&  +32.3~~&  2543.939&   +0.4~~ \\
\tableline
\end{tabular}
\\[5pt] 
{{$^{*}$\,Observed with original Cambridge spectrometer; weight {\small$^{1}\hskip-0.3em/_{\hskip-0.2em4}$}.\\[-3pt]
$^{\dagger}$\,Observed with Palomar spectrometer; weight {\small$^{1}\hskip-0.3em/_{\hskip-0.2em4}$}.\\[-3pt]
$^{\ddagger}$\,Observed with ESO `Coravel'; offset +0.8 km\,s$^{-1}$, weight 1.\\[-3pt]
$^{\S}$\,Observed with DAO spectrometer; weight {\small$^{1}\hskip-0.3em/_{\hskip-0.2em4}$}.\\[-3pt]
$^{\P}$\,Observed with Cambridge `Coravel'; offset $-$0.2 km\,s$^{-1}$, wt. 1.\\[-3pt]
$^{\parallel}$\,Observed by Strassmeier et al.~(2000); rejected (see text).
}}}
\end{table}

\begin{table}[h]\begin{center}
\setlength{\doublerulesep}{\arrayrulewidth}
\caption{Orbital elements for BD+39$\arcdeg$\,2587 \label{table-8}\vspace{2mm}}
\begin{tabular}{@{}ll@{}}
\hline\hline \\[-5pt]
\hskip1.5em$P$      &=~ 4.5938540~$\pm$~0.0000031 days$^{*}$\hskip 1.5 true cm \\
\hskip1.5em$T_{0}$ &=~ MJD 49438.0121~$\pm$~0.0013 \\
\hskip1.5em$\gamma$ &=~ --3.96~$\pm$~0.05 km\,s$^{-1}$ \\
\hskip1.5em$K$  &=~ 38.61~$\pm$~0.07 km\,s$^{-1}$  \\
\hskip1.5em$e$      &$\equiv$~ 0            \\
\hskip1.5em$\omega$ &is undefined in a circular orbit         \\
\hskip1.5em$a_{1}$\,sin\,$i$ &=~ 2.439~$\pm$~0.004 Gm \\ 
\hskip1.5em$f(m)$ &=~ 0.02747~$\pm$~0.00015 $M_{\odot}$ \\[5pt]
\multicolumn{2}{l}{~R.m.s. residual (wt. 1) = 0.42 km\,s$^{-1}$} \\[3pt]
\tableline
\end{tabular}
\\[5pt]\end{center}
{\small$^{*}$The true period, in the rest-frame of the system, is 4.593915~$\pm$~0.000003 days. It differs from the observed period  by 19 standard deviations.}
\end{table}

The orbit, whose period is determined to within about three millionths of a day (less than a third of a second) is very accurately circular: relaxation of the imposition of precise circularity in the solution reduces the weighted sum of squares of the residuals only from 14.50 to 14.48 (km\,s$^{-1})^2$. The velocities of star A have been adjusted and weighted on the same basis as those of B, and yield a mean of $-$3.95~$\pm$~0.07 km\,s$^{-1}$. The difference between the constant velocity of A and the $\gamma$-velocity of the B sub-system is therefore found to be only 0.01~$\pm$~0.09 km\,s$^{-1}$. 

\begin{table}[h!]
\setlength{\doublerulesep}{\arrayrulewidth}
{\footnotesize
\caption{Radial-velocity measurements of HD\,112733 \label{table-9}\vspace{2mm}}
\begin{tabular}{@{}rcrc@{}}
\tableline \\[-5pt]
UT Date~~~~\hspace{3mm} &    Velocity   & UT Date~~~ &    Velocity   \\
           &   km\,s$^{-1}$  &               &   km\,s$^{-1}$   \\[5pt]
\hline\hline\\[-5pt]
1973\hspace{2mm}Feb.\hspace{0.5em}25.16$^{*}$&$-$4.0~~ &1993\hspace{1.9mm}Mar.\hspace{0.4em}20.16~\,&$-$4.0~~\\
1977\hspace{2mm}May\hspace{0.5em}28.94$^{*}$&$-$4.9~~  &Dec.\hspace{.5em}28.23~\,&$-$3.5~~\\
1986\hspace{2mm}May\hspace{0.5em}27.91$^{*}$&$-$3.6~~  &1994\hspace{2mm}Apr.\hspace{.5em}30.04~\,&$-$4.0~~\\
 Nov.\hspace{0.4em}25.52$^{\dagger}$&$-$3.7~~           &1995\hspace{2.3mm}Jan.\hspace{1em}5.24~\,&$-$4.3~~\\
1987\hspace{2mm}Mar.\hspace{0.8em}4.13~\,&$-$3.9~~      &June\hspace{.8em}3.02~\,&$-$3.9~~\\
1991\hspace{2mm}Jan.\hspace{0.5em}30.07~\,&$-$4.0~~     &Dec.\hspace{.5em}27.19~\,&$-$4.3~~\\
 Feb.\hspace{1em}3.14~\,&$-$4.4~~                       &1996\hspace{2mm}Apr.\hspace{.9em}1.05~\,&$-$3.9~~\\
1992\hspace{2mm}Jan.\hspace{0.5em}21.21~\,&$-$4.1~~     &1997\hspace{2mm}May\hspace{1em}5.04$^{\S}$&$-$3.8~~\\
 Feb.\hspace{0.5em}28.49$^{\ddagger}$&$-$4.3~~          &2005\hspace{2mm}Jan.\hspace{.5em}23.22$^{\S}$&$-$3.7~~\\
 Apr.\hspace{0.5em}29.97~\,&$-$3.5~~                    &2007\hspace{2mm}Mar.\hspace{.3em}27.08$^{\S}$&$-$3.4~~\\
1993\hspace{2mm}Feb.\hspace{0.5em}15.13~\,&$-$4.2~~     &2009\hspace{2mm}May\hspace{.5em}26.92$^{\S}$&$-$3.8~~\\
\tableline
\end{tabular}
\\[5pt] }
{\footnotesize$^{*}$\,Observed with original Cambridge spectrometer.
\\[-3pt] $^{\dagger}$\,Observed with Palomar spectrometer.
\\[-3pt] $^{\ddagger}$\,Observed with ESO `Coravel'.
\\[-3pt] $^{\S}$\,Observed with Cambridge `Coravel'.
\\[-3pt] All others observed with Haute-Provence `Coravel'.}

\end{table}

Once again there is a stark conflict between the orbit given by GMFL and the one found here. GMFL refer to the ``very eccentric'' orbit, to which they attribute a period of 7.5656 days and an eccentricity of 0.3101, which will surely be difficult to maintain in the face of the evidence we show for a circular orbit with a period of about 4.6 days. GMFL, however, were apparently able to measure the spectroscopic secondary, which they imply to be of type M0\,V and is not detectable in our radial-velocity traces.

Despite the difference in the colours and (in the opposite sense) of the reported spectral types of the two stars, they give dips of very similar equivalent widths in radial-velocity traces, those of the B component being marginally the stronger. The mean $v$\,sin\,$i$ values for B are 6.2\,$\pm$\,0.4 km\,s$^{-1}$ according to the Haute-Provence traces and 5.4\,$\pm$\,0.7 km\,s$^{-1}$ according to the Cambridge ones. Taking 6 km\,s$^{-1}$ as a representative value, we obtain the star's projected radius, $R_{\star}$\,sin\,$i$, as about 0.54 $R_{\small\odot}$; possible distrust of its spectral classification increases the uncertainty as to its probable real radius, but it must surely be considerably larger than the projected one, so the inclination must be only moderate and there is no expectation of eclipses. The rotational velocity of the A component of the visual binary is too small to be determined accurately. What star \cite{favata95} observed, that had a $v$\,sin\,$i$ of 19 km\,s$^{-1}$, is a matter for conjecture.

\subsection{HD~138157 (OX Ser)}

Unlike the other stars discussed in this paper, OX\,Ser is a giant. It has a V magnitude given by \citet{cl93} as $7\textsuperscript{m}.145$ but shown to be slightly variable by \textit{Hipparcos}, which found its parallax to be 5.07~$\pm$~1.00 milliseconds of arc, indicating an absolute visual magnitude of about +$0\textsuperscript{m}.6$~$\pm$~$0\textsuperscript{m}.4$. The spectral type has been given by GMFL as K0\,III, which is quite consonant with the colour indices of (B\,--\,V) $1\textsuperscript{m}.055$, (U\,--\,B) $0\textsuperscript{m}.718$ \citep{cl93}.

\begin{figure}[h]
\begin{center}
\includegraphics[scale=0.35, angle=-90]{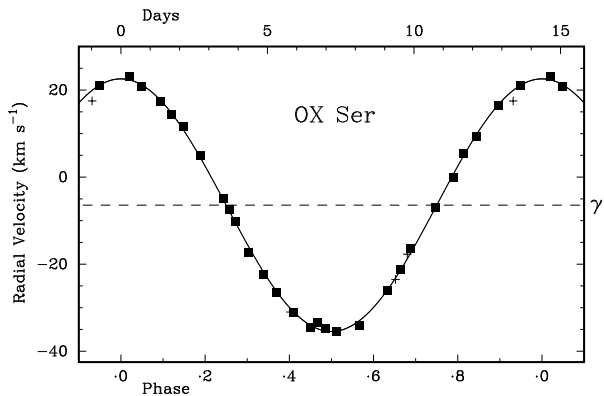}
\caption[] {The observed radial velocities of OX\,Serpentis plotted as a function of phase, with the velocity
curve corresponding to the adopted orbital elements drawn through them. The Cambridge data are plotted
as filled squares and the \citet{strass00} ones as plusses, as usual.\label{figure-5}}
\end{center}
\end{figure}

The 27 Cambridge radial-velocity measurements are set out in Table~\ref{table-10}, along with the four offered by \citet{strass00}. Alone, the recent velocities give a circular orbit with a period of 14.365~$\pm$~0.004 days. The Strassmeier measurements are about 270 cycles earlier, at which the 1-$\sigma$ phasing uncertainty is little more than 1 day, so the cycle count is quite secure. Addition of the early observations, with an offset of +0.8 km\,s$^{-1}$ and a weighting of  {\small$^{1}\hskip-0.3em/_{\hskip-0.2em4}$} for comparability of variances, improves the precision of the period by a factor of more than ten. The final orbit is plotted in Fig.\,\ref{figure-5} and its elements are shown in Table~\ref{table-11}. The adopted orbit is exactly circular. If $e$ and $\omega$ are allowed as free parameters, the slight eccentricity that is found is non-significant, as is demonstrated by the use of Bassett's (1978) second statistical test that compares the sums of squares of the residuals obtained with $e$ free and with $e$ fixed at 0; the sums are 11.27 and 11.79 (km\,s$^{-1})^{2}$ respectively. The only information that GMFL gave about their orbit for the star is the period, which in this case is in agreement with the one that we find. The photometric period has also been found to be 14.3 days, by \citet{strass00}, after \textit{Hipparcos} (which was responsible for the initial discovery of the variability) had proposed 7.1853 days, very close to half the orbital period.

\begin{table}[h!!]
\setlength{\doublerulesep}{\arrayrulewidth}
{\footnotesize
\caption{Radial-velocity observations of OX\,Ser \label{table-10}}
{Except as noted, the observations were made at Cambridge}\\[5pt]
\begin{tabular}{@{}rrrrr@{}}
\tableline \\[-5pt]
Date (UT)~~~~ &        MJD~~~ &   Velocity &      Phase~~&    (O -- C) \\
           &            &     km\,s$^{-1}$ &            &     km\,s$^{-1}$ \\[5pt]
\hline\hline \\[-5pt]
 1998\hspace{0.5em}Apr.\hspace{.5em}20.40$^{*}$&  50923.40&  $-$23.5~~&$\overline{277}$.652&  $-$0.3~~~\\
         Sept.\hspace{0.2em}15.10$^{*}$&  51071.10&  +17.5~~&$\overline{267}$.932&  $-$2.4~~~\\
&&&&\\
 1999\hspace{0.5em}Feb.\hspace{.5em}12.55$^{*}$&  51221.55&  $-$31.0~~&$\overline{256}$.403&  $-$0.8~~~\\
         16.54$^{*}$&  225.54&  $-$17.7~~&.680&  +1.1~~~\\
&&&&\\
2009\hspace{.5em}Mar.\hspace{1em}6.22~~&  54896.22&  +11.6~~&     0.148&   +0.8~~~\\
                 21.17~~&    911.17&   +4.9~~&     1.189&   +0.5~~~\\
                 30.15~~&    920.15&   +5.4~~&      .814&   +0.5~~~\\
Apr.\hspace{1.1em}1.10~~&    922.10&  +21.1~~&      .950&    0.0~~~\\
                  2.11~~&    923.11&  +23.1~~&     2.020&   +0.8~~~\\
                 21.06~~&    942.06&  $-$22.3~~&     3.339&   $-$0.5~~~\\
                 22.08~~&    943.08&  $-$31.0~~&      .410&   $-$0.1~~~\\
                 29.07~~&    950.07&  +16.4~~&      .896&   $-$0.2~~~\\
   May\hspace{1em}4.07~~&    955.07&   $-$5.0~~&     4.244&   +0.4~~~\\
                  7.03~~&    958.03&  $-$34.6~~&      .450&   $-$0.5~~~\\
                 23.06~~&    974.06&  $-$34.1~~&     5.566&   $-$1.1~~~\\
                 24.03~~&    975.03&  $-$26.0~~&      .633&   $-$0.2~~~\\
                 27.05~~&    978.05&   +9.4~~&      .844&   $-$0.3~~~\\
                 29.99~~&    980.99&  +20.8~~&     6.048&   $-$0.4~~~\\
                 31.02~~&    982.02&  +14.3~~&      .120&   $-$0.4~~~\\
 June\hspace{.8em}2.00~~&    984.00&   $-$7.4~~&      .258&   +0.5~~~\\
                 17.03~~&    999.03&  $-$17.3~~&     7.304&   $-$1.2~~~\\
                 17.99~~&    999.99&  $-$26.6~~&      .371&   $-$0.2~~~\\
                 20.01~~&  55002.01&  $-$35.5~~&      .511&   $-$0.1~~~\\
                 24.00~~&    006.00&    0.0~~&      .789&   $-$0.6~~~\\
                 30.94~~&    012.94&  $-$10.1~~&     8.272&   +0.3~~~\\
  July\hspace{1em}4.00~~&    016.00&  $-$34.9~~&      .485&   +0.4~~~\\
                  6.92~~&    018.92&  $-$16.4~~&      .688&   +1.1~~~\\
                 20.95~~&    032.95&  $-$21.3~~&     9.664&    0.0~~~\\
Aug.\hspace{.5em}15.86~~&    058.86&  $-$33.4~~&    11.468&   +1.5~~~\\
                 19.86~~&    062.86&   $-$7.0~~&      .746&   +0.1~~~\\
                 24.87~~&    067.87&  +17.3~~&    12.095&   $-$0.3~~~\\[3pt]
\tableline                                   
\end{tabular}
\\[5pt] $^{*}$Observed by \citet{strass00}; weight {\small$^{1}\hskip-0.3em/_{\hskip-0.2em4}$}}.
\end{table}

\begin{table}[h!!]\begin{center}
\setlength{\doublerulesep}{\arrayrulewidth}
\caption{Orbital elements for OX\,Serpentis \label{table-11}\vspace{2mm}}
\begin{tabular}{@{}ll@{}}
\hline\hline\\[-5pt]
\hskip1.5em$P$      &=~ 14.36843~$\pm$~0.00028 days\hskip 1.5 true cm \\
\hskip1.5em$T_{0}$ &=~ MJD 54836.613~$\pm$~0.014 \\
\hskip1.5em$\gamma$ &=~ $-$6.44~$\pm$~0.13 km\,s$^{-1}$ \\
\hskip1.5em$K$  &=~ 29.01~$\pm$~0.18 km\,s$^{-1}$  \\
\hskip1.5em$e$      &$\equiv$~ 0            \\
\hskip1.5em$\omega$ &is undefined in a circular orbit         \\
\hskip1.5em$a_{1}$\,sin\,$i$ &=~ 5.73~$\pm$~0.04 Gm \\ 
\hskip1.5em$f(m)$ &=~ 0.0364~$\pm$~0.0007 $M_{\odot}$ \\[5pt]
\multicolumn{2}{l}{~R.m.s. residual (wt. 1) = 0.62 km\,s$^{-1}$} \\[3pt]
\hline
\end{tabular}
\end{center}\end{table}

The brightness of OX\,Ser in comparison with the other stars treated here enables radial-velocity traces of reasonably good S/N to be obtained without exorbitant integration times, which is helpful because the dips to be measured are particularly wide and shallow. The mean rotational velocity is 33.3~$\pm$~0.3 km\,s$^{-1}$; it indicates a projected stellar radius of 9.5 $R_{\odot}$.

\subsection{HD~143705}

\begin{figure}[h]
\begin{center}
\includegraphics[scale=0.35, angle=-90]{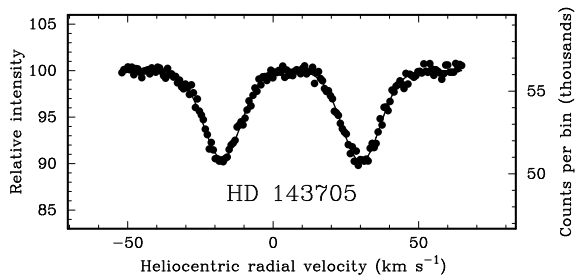}
\caption[] {Radial-velocity trace of HD~143705, obtained with the Cambridge \textit{Coravel} on 2009 July 4, illustrating the well separated double lines.\label{figure-6}}
\end{center}
\end{figure}

This object, like HD\,89959, is double-lined and has components that are almost identical with one another. A radial-velocity trace illustrating those facts appears as Fig.\,\ref{figure-6}. In the post-war years the star was on the David Dunlap Observatory (DDO) programme on late-type stars in the +25$\arcdeg$--30$\arcdeg$ zone of declination \citep{heard}, in the course of which it was classified G0\,V and measured five times for radial velocity. Remarkably, it was identified as a spectroscopic binary, and on that account the five observations were published individually, with dates, instead of simply as a mean. We say ``remarkably'' because the resolution that was brought to bear was not enough for the double lines to be recognized, and when measured as single-lined the spectrum ought to have shown altogether inappreciable variations. The five velocities show no relationship to orbital phase. In fact four of them agree with one another and with the $\gamma$-velocity within the uncertainties of the observations; there is just one outlier. We recall that a few errors have been detected previously \citep{griffin79} in the DDO results.

The magnitude and colours of HD 143705 have been determined by \citet{eggen}, \citet{priser} and \citet{oja}, with accordant results close to V~=~$7\textsuperscript{m}.97$, (B\,$-$\,V)~=~$0\textsuperscript{m}.60$, (U\,$-$\,B)~=~$0\textsuperscript{m}.08$. The \textit{Hipparcos} parallax of $0\arcsec.01616~\pm~0\arcsec.00098$ translates to a distance modulus of $3\textsuperscript{m}.96~\pm~0\textsuperscript{m}.13$ and so to an absolute magnitude of $4\textsuperscript{m}.01$ with the same uncertainty. Since the system consists of two almost equal stars, each of the components must have an absolute magnitude near to $4\textsuperscript{m}.76$. It is seen that both the luminosities and the colours support the DDO early-G main-sequence classification of the system. There is a very small, but distinct, difference between the components. Anticipating a result from Table~\ref{table-13}, we find that there is a difference in mass of $1.3~\pm~0.4$ per cent, which corresponds to about half a spectral sub-type. The areas of the dips seen in radial-velocity traces, which have mean values of $1.68~\pm~0.02$ and and $1.60~\pm~0.02$~km\,s$^{-1}$ for the respective components, differ by about $5~\pm~2$ per cent, equivalent to $0\textsuperscript{m}.05~\pm~0\textsuperscript{m}.02$ in luminosity terms or to about a third of a sub-type, in the same sense as the mass difference. Mean projected rotational velocities of about 3~km\,s$^{-1}$ are found for both components but are not very accurately established. The equatorial velocity of a star of solar radius rotating at a rate pseudo-synchronized with the orbital revolution in HD~143705 would be about 3.9~km\,s$^{-1}$, but there is no strong presumption of synchronization at the 14.3-day period of the system.

\begin{figure}[h]
\begin{center}
\includegraphics[scale=0.35, angle=-90]{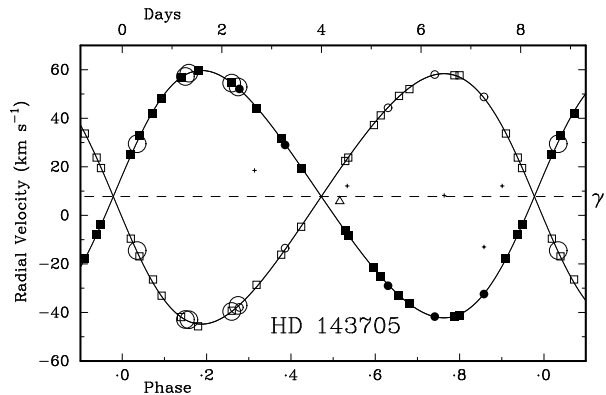}
\caption[] {The observed radial velocities of HD~143705 plotted as a function of phase, with the velocity curves
corresponding to the adopted orbital elements drawn through them. The same conventions as in Fig.\,\ref{figure-3} are
adopted for the plotting symbols. In addition, the five early photographic observations (Heard 1956), which
were not used in the solution of the orbit, appear as plusses. We have attempted to retrieve the measurements
that were made by \citet{galvez06} but were not published except in a graph plotted on an erroneous
period; they are shown here, for both components, as large open circles, but were not utilized in the orbital
solution.\label{figure-7}}
\end{center}
\end{figure}

There are 21 recent Cambridge radial-velocity observations, which are listed in Table~\ref{table-12}; the DDO ones, and six obtained on consecutive nights (at times specified to $10^{-8}$ day---better than a millisecond---but here rounded to only two decimals) by \citet{wsh03}, are included at the head of the table. The Cambridge data produce an orbit with a significant eccentricity and a period of 8.4687\,$\pm$\,0.0004 days. In the $\sim$440 cycles back to the \citeauthor{wsh03} measurements the uncertainty in the phasing increases only to about 0.2 days, so it does not permit any ambiguity in the cycle count, and those measurements can be brought in to refine the orbit. For them to do that, the identities of the components need to be inverted from the designations given by \citeauthor{wsh03}, but that calls for no explanation since the components are mutually indistinguishable. As in the case of the HD~89959 velocities from the same source, an empirically determined zero-point adjustment of +1.5~km\,s$^{-1}$ has been found desirable for them, but in this case they deserve a weighting equal to that of the recent data. The resulting orbit is portrayed in Fig.\,\ref{figure-7} and its elements appear in Table~\ref{table-13}.

\begin{table*}\clearpage
\setlength{\doublerulesep}{\arrayrulewidth}
\setlength\tabcolsep{1.2mm}
\footnotesize
\caption{Radial-velocity observations of HD\,143705 \label{table-12}}
{Except as noted, the observations were made at Cambridge}\\[5pt]
\begin{tabular}{@{}rrrrrrr@{}}
\hline\hline\\[-5pt]
Date (UT)~~~~ & MJD~~ & \multicolumn{2}{c}{Velocity} & Phase &\multicolumn{2}{c}{(O -- C)} \\
      &      & Prim.~~&  Sec.~~~&       &  Prim.~~&  Sec.~~~\\
      &      & km\,s$^{-1}$ &  km\,s$^{-1}$ &       &  km\,s$^{-1}$ &  km\,s$^{-1}$ \\[5pt]
\tableline\\[-5pt]
 1948\hspace{0.5em}June\hspace{0.4em}20.17$^{*}$&  32722.17& \multicolumn{2}{c}{+8.2} &$\overline{2615}$.764&---\hspace{0.5em}~~&---\hspace{0.5em}~~~\\
&&&&&\\
 1949\hspace{0.5em}May\hspace{0.5em}17.25$^{*}$&  33053.25& \multicolumn{2}{c}{--13.0} &$\overline{2576}$.858&---\hspace{0.5em}~~&---\hspace{0.5em}~~~\\
&&&&&\\
 1950\hspace{0.5em}May\hspace{1em}3.32$^{*}$&  33404.32&  \multicolumn{2}{c}{+18.5} &$\overline{2534}$.313&---\hspace{0.5em}~~&---\hspace{0.5em}~~~\\
           8.30$^{*}$&    409.30&   \multicolumn{2}{c}{+12.1}&.901&---\hspace{0.5em}~~&---\hspace{0.5em}~~~\\
&&&&&~~~\\
 1951\hspace{0.5em}June\hspace{0.4em}24.15$^{*}$&  33821.15&  \multicolumn{2}{c}{+12.1}&$\overline{2485}$.534&---\hspace{0.5em}~~&---\hspace{0.5em}~~~\\
&&&&&\\
 1999\hspace{0.5em}Mar.\hspace{0.5em}27.47$^{\dagger}$&  51264.47&    +52.1&    $-$37.9&$\overline{425}$.277&    +0.2~~&    $-$1.0~~~\\
           28.39$^{\dagger}$&    265.39&    +29.0&    $-$13.5&.386&    $-$0.3~~&    +0.6~~~\\
           29.49$^{\dagger}$&    266.49&      +5.8&.516&---\hspace{0.5em}&---\hspace{0.5em}~~~\\
           30.46$^{\dagger}$&    267.46&    $-$29.0&    +44.3&.630&    $-$0.5~~&    $-$0.2~~~\\
           31.40$^{\dagger}$&    268.40&    $-$41.7&    +58.0&.741&    +0.1~~&    +0.1~~~\\
  Apr.\hspace{1em}1.39$^{\dagger}$&    269.39&    $-$32.4&    +48.8&.858&    $-$0.1~~&    +0.4~~~\\
&&&&&\\
 2004\hspace{0.5em}Apr.\hspace{1em}5.12$^{\ddagger}$&  53100.12&    +29.5&    $-$14.5&$\overline{  208}$.035&    $-$0.8~~&    +0.6~~~\\
           6.09$^{\ddagger}$&    101.09&    +57.2&    $-$42.9&.149&    $-$0.7~~&    +0.1~~~\\
           6.16$^{\ddagger}$&    101.16&    +58.6&    $-$43.0&.157&     0.0~~&    +0.7~~~\\
           7.02$^{\ddagger}$&    102.02&    +54.5&    $-$39.6&.259&    $-$0.1~~&    0.0~~~\\
           7.15$^{\ddagger}$&    102.15&    +52.9&    $-$37.1&.274&    +0.5~~&    +0.3~~~\\
&&&&&\\
 2009\hspace{0.5em}Feb.\hspace{1.1em}7.26~\,&  54869.26&     $-$7.8&    +23.8&     0.938&    +0.1~~&    +0.1~~~\\
           12.26~\,&    874.26&     $-$6.4&    +22.4&     1.529&    0.0~~&    +0.3~~~\\
  Mar.\hspace{0.4em}27.18~\,&    917.18&    $-$21.6&    +37.2&     6.597&    +0.3~~&    $-$0.6~~~\\
           30.16~\,&    920.16&     $-$3.7&    +19.5&.949&    +0.3~~&    $-$0.1~~~\\
  Apr.\hspace{1em}1.11~\,&    922.11&    +59.7&    $-$45.7&     7.179&    +0.1~~&    $-$1.0~~~\\
           21.07~\,&    942.07&     $-$8.4&    +23.8&     9.536&    $-$0.3~~&    $-$0.1~~~\\
           22.09~\,&    943.09&    $-$33.0&    +49.2&.656&    $-$0.1~~&    +0.2~~~\\
  May\hspace{1em}4.08~\,&    955.08&    +41.9&    $-$26.4&    11.072&    $-$1.0~~&    +1.4~~~\\
            7.06~\,&    958.06&    +19.5&     $-$4.7&.424&    $-$0.4~~&    $-$0.2~~~\\
           23.10~\,&    974.10&    +44.1&    $-$28.6&    13.318&    $-$0.4~~&    +0.9~~~\\
           27.08~\,&    978.08&    $-$41.6&    +57.7&.788&    0.0~~&    0.0~~~\\
           29.04~\,&    980.04&    +25.0&     $-$9.6&    14.020&    +0.4~~&    $-$0.3~~~\\
           30.05~\,&    981.05&    +56.6&    $-$41.9&.139&    $-$0.2~~&     0.0~~~\\
           31.07~\,&    982.07&    +54.9&    $-$39.3&.259&    +0.4~~&    +0.3~~~\\
  June\hspace{0.3em}30.99~\,&  55012.99&    $-$17.8&    +33.7&    17.910&    +0.1~~&   $-$0.1~~~\\
  July\hspace{1em}4.94~\,&    016.94&    +31.5&    $-$16.2&    18.377&     0.0~~&    +0.1~~~\\
           6.94~\,&    018.94&    $-$25.2&    +41.2&.613&     0.0~~&     0.1~~~\\
          27.94~\,&    039.94&    +48.1&  $-$33.1&    21.093&    -0.3~~&    +0.3~~~\\
Aug.\hspace{0.5em}18.86~\,&    061.86&  $-$36.3&    +52.0&    23.681&    +0.2~~&    -0.6~~~\\
          19.87~\,&    062.87&  $-$41.4&    +57.7&  .800&    -0.6~~&    +0.8~~~\\
          21.90~\,&    064.90&    +32.9&  $-$16.9&    24.040&    +0.6~~&    +0.1~~~\\[3pt]
\tableline
\end{tabular}
\\[5pt] $^{*}$\,Observed by  \citet{heard}; weight 0.\\
$^{\dagger}$\,Observed by  \citet{wsh03}; weight 1.\\
$^{\ddagger}$\,`Reclaimed' observation by  \citet{galvez06} (see text); weight 0.\\
\end{table*}

\citeauthor{wsh03}, while recognizing that their observations did not cover as much as a single cycle, hazarded two attempts to obtain a solution from them. In one, they used all six observations, five of which were double-lined, and in the other they discarded the one measure that was of the blend when the velocities of the two components were too close to the $\gamma$-velocity for the spectra to be resolved. The second could be expected to be the more realistic one; the period was in that case found to be 8.43 days, very near the true value, but its uncertainty was nearly a whole day, and the values of $e$ and $\omega$ were likewise very uncertain. 

Subsequently, GMFL utilized the Wichmann,\break Schmitt, and Hubrig observations in conjunction with five of their own to determine the orbit anew. The only elements that they gave for it were the period and the eccentricity, which they put at 8.1130 days and 0.1267, respectively. The former is obviously in conflict with our value of 8.46868 days, which has an uncertainty of less than 2 units in the fifth decimal (less than two seconds of time). Knowing the date of their observing run, we can reconstruct (just as for HD~89959) the manner in which GMFL must have guessed at the number of cycles that had elapsed between the \citeauthor{wsh03} epoch and their own, since they had no means of ascertaining what it really was; the number that they guessed was evidently 226, whereas the true number was 216{\small$^{1}\hskip-0.3em/_{\hskip-0.2em2}$}, the odd half-cycle arising from the unrecognized inversion of the identities of the components with respect to the \citeauthor{wsh03} choice. 

It is easily possible to go much further in reconstructing the GMFL material. The radial velocities, and the relative dates of the observations, have been read back from an enlarged copy of their graph of the orbit of HD~143705---a procedure that has worked well in previous instances, e.g.\,that of $\gamma$ Cephei \citep{griffin92}, notwithstanding that it was regarded with unconcealed astonishment \citep{hatzes} by the authors whose data were thereby retrieved! The observations were made on three consecutive nights. They fit very nicely to one lobe of the velocity curves but not to the other, so there is only one set of times that fits in each 8{\small$^{1}\hskip-0.3em/_{\hskip-0.2em2}$}-day orbital cycle. Within the single GMFL observing run, 2004 March 27 to April 7, there is at first sight an ambiguity, but it is easily resolved by the consideration that HD~143705 would have been on the local meridian at about $3\textsuperscript{h}$ UT. The relevant nights are thereby identified as those starting on 2004 April 4--6. By suitably adjusting the zero-point of the time scale and by adoption of an empirical offset of +1.5 km\,s$^{-1}$ to the radial-velocity zero-point, the GFML data can be steered, as a monolithic block in which the ten data points are mutually fixed, into excellent agreement (see Fig.\,\ref{figure-7}) with the orbit that we have determined from the properly published observations plus the new Cambridge ones presented here. Inasmuch as the need arbitrarily to choose time and velocity zero-points has cost only two degrees of freedom, eight degrees remain in the GMFL data, so it could be argued quite validly that they could still perfectly well be utilized in the determination of the orbit, especially since their residuals are scarcely worse than those of our own observations. That logical position may not appeal to everyone, and as we have no real need of additional data to determine the orbit we forego the actual use of the GMFL observations, contenting ourselves with illustrating in Fig.\,\ref{figure-7} where they fall in the orbit graph. Nevertheless we have included our reconstruction of the times and velocities in Table~\ref{table-12}.

\begin{table}[h!]
\setlength{\doublerulesep}{\arrayrulewidth}
\begin{center}
\caption{Orbital elements of HD~143705 \label{table-13}\vspace{2mm}}
\begin{tabular}{@{}ll@{}}
\hline\hline\\[-5pt]
\hskip1.5em$~P$      &=~ 8.468685~$\pm$~0.000017 days$^{*}$ \\
\hskip1.5em$~T_{}$ &=~ MJD 54260.036~$\pm$~0.022 \\
\hskip1.5em$~\gamma$ &=~ +7.77~$\pm$~0.07 km\,s$^{-1}$ \\
\hskip1.5em$~K_{1}$  &=~ 50.95~$\pm$~0.14 km\,s$^{-1}$  \\
\hskip1.5em$~K_{2}$  &=~ 51.59~$\pm$~0.14 km\,s$^{-1}$  \\
\hskip1.5em$~q$      &=~ 1.013~$\pm$~0.004 (=$m_{1}/m_{2}$) \\
\hskip1.5em$~e$      &=~ 0.1205~$\pm$~0.0017             \\
\hskip1.5em$~\omega$ &=~ 279.1~$\pm$~1.0 degrees         \\
\hskip1.5em$~a_{1}$\,sin\,$i$ &=~ 5.889~$\pm$~0.016 Gm \\ 
\hskip1.5em$~a_{2}$\,sin\,$i$ &=~ 5.963~$\pm$~0.016 Gm \\ 
\hskip1.5em$~f(m_{1})$ &=~ 0.1138~$\pm$~0.0009 $M_{\odot}$ \\
\hskip1.5em$~f(m_{2})$ &=~ 0.1181~$\pm$~0.0009 $M_{\odot}$ \\
\hskip1.5em$~m_{1}$\,sin$^{3}\,i$ &=~ 0.467~$\pm$~0.003 $M_{\odot}$ \\
\hskip1.5em$~m_{2}$\,sin$^{3}\,i$ &=~ 0.461~$\pm$~0.003 $M_{\odot}$ \\[5pt]
\multicolumn{2}{l}{R.m.s. residual (unit weight) = 0.44 km\,s$^{-1}$} \\[3pt]
\hline
\end{tabular}\\[5pt]\end{center}
$^{*}${\small The true period, in the rest-frame of the system, is 8.468465~$\pm$~0.000018 days. It differs from the observed period  by 12 standard deviations.}
\end{table}

Analogous reconstructions could be pursued, if that seemed desirable, for some of the others stars treated by GMFL; but how much better it would have been if those authors had published their data in the first place! If HD~143705 is a representative example, the data themselves are excellent: it is the discussions that are open to objection.

\subsection{HD\,160934}

HD\,160934 is considerably fainter than the other stars treated here, at V $\sim$ $10\textsuperscript{m}.28$ \citep{weis,esa}, and is variable over a range of $0\textsuperscript{m}.1$ or so. In view of its faintness it is surprising that it features in the Henry Draper Catalogue \citep{cp22}, in which its type is given as Ma. \citet{vyss} gave its type as K8\,V, and subsequent authors have either quoted that or proposed very similar classifications. \citet{henry} reported photometric variations in either 1.842 or 2.181 days (aliases of one another, equidistant on either side of 2 days when expressed as frequencies), but \textit{Hipparcos} (vol. 11, p. PN84) specifically notes that it can offer no confirmation---referring to the period, not the fact of variability. \citet{pandey} proposed a period of about 43 days, but they had in effect only five independent data points, which could be consonant with many quite different periods; we also have reservations about those authors' assessments of standard errors.

The star rotates rapidly, has been observed as a far-ultraviolet \citep{pounds} and X-ray \citep{hun} source, shows H${\alpha}$ in strong emission and has a considerable Li I $\lambda$6708-\AA\ line \citep{mullis}, and in general exhibits every sign of being a young and chromospherically active star, even though it does not appear in the recent CABS3.

HD~160934 has been suggested as a member of the `local association' \citep{montes} and of the `AB Dor moving group' \citep{zsb04,lopez}, but not with much enthusiasm---which is hardly surprising since without a reliable radial velocity its kinematics are indeterminate.

\citet{weis} measured, at $14\textsuperscript{m}.7$, a visual companion star {20\arcsec} from HD~160934; it has seemed increasingly likely that it shares the proper motion of the principal star and thus may be a physical companion. It is clearly seen in the pictures that Shara and collaborators \citep{ssm,shara} repeatedly published as finding charts. A discovery of particular significance to our present interest was made by \citet{hormuth} and promptly confirmed by \citet{laf}: HD~160934 is a close visual binary, with a separation of about {0\arcsec}.21. The former group found a $\Delta m$ of about $1\textsuperscript{m}.2$~$\pm$~$0\textsuperscript{m}.15$ at $\lambda$ $\sim$ 8300 \AA; the latter authors gave a $\Delta m$ of about $0\textsuperscript{m}.85$ at a wavelength of 1.6 $\mu$, thereby indicating that the companion is significantly redder even than the primary star. \citeauthor{hormuth}, after discussing the parallax, colours, and stellar models, opted for masses of 0.69 and 0.57 $M_{\odot}$ for the pair, and spectral types near K5 and M0\,V.

Before the paper by GMFL, there seem to have been only two radial velocities for HD~160934 in the literature, and neither of them was tied to a particular date. \citet{henry} considered the velocity ``const:'' on the basis of three spectrograms which they describe as having been taken ``recently'' (their paper was submitted on 1995 August 9); their result was $-26.7\,\pm\,0.1$~km\,s$^{-1}$, and the apparent constancy of the velocity, even over a matter of days, at least seemed to show that the star was not involved in a binary system having the photometric period of about 2 days that they observed at that time. They noted, however, quite rapid rotation, with $v$\,sin\,$i$~=~13\,$\pm$\,1\ km\,s$^{-1}$, subsequently altered by \citet{fekel97}, who still described the radial velocity as constant, to 16.4 km\,s$^{-1}$. \citet{zsb04} listed a radial velocity of $-35.6\,\pm\,0.7$ km\,s$^{-1}$ for HD~160934, but they too omitted to give a date for it; they reported that their observing campaign began in 2001, so all we can say is that it must have been in the interval 2001--2004. During that time we believe that the star passed that velocity twice, first on the descending branch of the velocity curve and then on the ascending one. They gave a $v$\,sin\,$i$ of 17 km\,s$^{-1}$.

\begin{figure}[h]
\begin{center}
\includegraphics[scale=0.35, angle=-90]{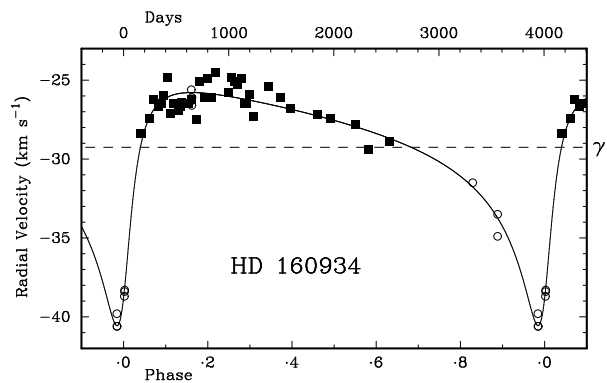}
\caption[] {The observed radial velocities of HD\,160934 plotted as a function of phase, with the velocity curve
corresponding to the adopted orbital elements drawn through them. Once again the Cambridge data are
plotted as filled squares; the open circles are measurements kindly provided privately by Dr.\,D.\,Montes. The
orbit evidently must remain preliminary until another periastron passage has taken place and been carefully
observed.\label{figure-8}}
\end{center}
\end{figure}

GMFL showed for HD\,160934 a radial-velocity curve having a long orbital period and a high eccentricity; the star had not by any means been seen round a complete cycle. Most of the plotted data were actually provided to Dr.\,D.\ Montes (the 'M' of GMFL) by one of the writers, who has been observing the star since 2002 (Table~\ref{table-14}), but who did not realize that the velocities were to be used in the premature orbit publication. It is, however, fair to acknowledge that it was Dr.\,Montes who first divined the nature of the orbit, at a time when the present authors' observations appeared to show only a quasi-constant velocity. The GMFL orbit has a period of 6246.2318 days; despite the 8-digit precision with which it is asserted it is certainly in error in the very first digit---by thousands of days! Without wishing to suggest that it is anything but another very preliminary orbit, with formally computed standard errors that are almost certainly too optimistic, we present in Table~\ref{table-15} the elements of our own orbit, which is plotted in Fig.$\,$\ref{figure-8}. It must be made clear that the calculation of the orbit involves, and indeed is dependent upon, a small number of observations which we do not consider ourselves at liberty to publish numerically because they were made by GMFL. Our radial-velocity traces yield a mean $v$\,sin\,$i$ value of 16.4~$\pm$~0.3 km\,s$^{-1}$.

\begin{table}
\setlength{\doublerulesep}{\arrayrulewidth}
\setlength\tabcolsep{1.5mm}
{\footnotesize
\caption{Radial-velocity observations of HD~160934 \label{table-14}\vspace{2mm}}
\begin{tabular}{@{}rrrrrr@{}}
\tableline \\[-5pt]
Date (UT)~~~ &        MJD~~~~&   Velocity &      Phase &    (O -- C) \\
           &            &     km\,s$^{-1}$ &            &     km\,s$^{-1}$ \\[5pt]
\hline\hline\\[-5pt]

 2002\hspace{0.5em}Dec.\hspace{1em}4.76&  52612.76&  $-$28.4~~&  0.040&   +0.8~~\\
                     17.72&    625.72&  $-$28.4~~&.043&   +0.4~~\\
&&&&&~~\\
 ~~2003\hspace{0.5em}Mar.\hspace{0.9em}3.23&  52701.23&  $-$27.4~~&  0.062&   $-$0.1~~\\
      Apr.\hspace{1em}8.14&    737.14&  $-$26.2~~&.071&   +0.7~~\\
    May\hspace{0.5em}29.01&    788.01&  $-$26.7~~&.084&   $-$0.2~~\\
   June\hspace{0.4em}25.00&    815.00&  $-$26.5~~&.091&   $-$0.2~~\\
   July\hspace{0.6em}13.95&    833.95&  $-$26.0~~&.095&   +0.2~~\\
   Aug.\hspace{0.4em}14.95&    865.95&  $-$24.8~~&.103&   +1.3~~\\
  Sept.\hspace{0.2em}13.88&    895.88&  $-$27.1~~&.111&   $-$1.1~~\\
   Oct.\hspace{0.5em}11.89&    923.89&  $-$26.5~~&.118&   $-$0.6~~\\
   Nov.\hspace{0.4em}27.74&    970.74&  $-$26.9~~&.129&   $-$1.0~~\\
   Dec.\hspace{0.5em}15.72&    988.72&  $-$26.7~~&.134&   $-$0.9~~\\
                     28.72&  53001.72&  $-$26.4~~&.137&   $-$0.6~~\\
&&&&&~~\\ 
 2004\hspace{0.5em}Mar.\hspace{0.8em}2.23&  53066.23&  $-$26.5~~&  0.153&   $-$0.7~~\\
                      3.18&    098.18&  $-$26.2~~&.161&   $-$0.4~~\\
    May\hspace{0.5em}19.09&    144.09&  $-$27.5~~&.173&   $-$1.7~~\\
   June\hspace{0.4em}15.05&    171.05&  $-$25.1~~&.179&   +0.7~~\\
   Aug.\hspace{0.9em}8.00&    225.00&  $-$26.1~~&.193&   $-$0.3~~\\
  Sept.\hspace{0.7em}1.94&    249.94&  $-$24.9~~&.199&   +0.9~~\\
   Oct.\hspace{1em}5.85&    283.85&  $-$26.1~~&.208&   $-$0.2~~\\
   Nov.\hspace{0.4em}13.80&    322.80&  $-$24.5~~&.217&   +1.4~~\\
&&&&&~~\\ 
 2005\hspace{0.5em}Mar.\hspace{0.3em}25.17&  53454.17&  $-$25.8~~&  0.250&   +0.2~~\\
   Apr.\hspace{0.5em}22.12&    482.12&  $-$24.8~~&.257&   +1.3~~\\
   May\hspace{0.5em}15.07&    505.07&  $-$25.1~~&.263&   +1.0~~\\
   June\hspace{0.4em}11.03&    532.03&  $-$25.3~~&.270&   +0.8~~\\
   July\hspace{0.5em}21.98&    572.98&  $-$24.9~~&.280&   +1.3~~\\
   Aug.\hspace{0.4em}15.98&    597.98&  $-$26.5~~&.286&   $-$0.3~~\\
   Sept.\hspace{0.7em}8.94&    621.94&  $-$26.5~~&.292&   $-$0.3~~\\
   Oct.\hspace{1em}4.83&    647.83&  $-$25.9~~&.299&   +0.4~~\\
   Nov.\hspace{0.4em}16.75&    690.75&  $-$27.3~~&.309&   $-$1.0~~\\
&&&&&~~\\ 
 2006\hspace{0.5em}Apr.\hspace{1em}4.16&  53829.16&  $-$25.4~~&  0.344&   +1.1~~\\
   Aug.\hspace{0.9em}2.02&    949.02&  $-$26.1~~&.374&   +0.6~~\\
   Nov.\hspace{0.9em}1.82&  54040.82&  $-$26.8~~&.397&    0.0~~\\
&&&&&~~\\ 
 2007\hspace{0.5em}July\hspace{0.5em}19.01&  54300.01&  $-$27.2~~&  0.461&   +0.1~~\\
   Nov.\hspace{0.4em}15.77&    419.77&  $-$27.4~~&.491&   +0.1~~\\
&&&&&~~\\ 
 2008\hspace{0.5em}July\hspace{0.5em}13.06&  54660.06&  $-$27.8~~&  0.551&   +0.2~~\\
   Nov.\hspace{0.9em}7.80&    777.80&  $-$29.4~~&.581&   $-$1.2~~\\
&&&&&~~\\ 
 2009\hspace{0.5em}May\hspace{0.5em}29.08&  54980.08&  $-$28.9~~&  0.631&   $-$0.2~~\\[3pt]
\tableline
\end{tabular}}
\end{table}

\citet{hormuth} not only detected the companion themselves on 2006 July 8 but retrieved an HST observation dating from 1998 June 30 and also showing it. In the HST observation the companion was closer ({0\arcsec}.155, against {0\arcsec}.215 in 2006), and its position angle was larger by $4\arcdeg.6$~$\pm$~$0\arcdeg.4$. That is to say, after an interval of eight years there had been only a rather small change in position angle and a moderate change in angular separation. It is clearly of interest to discuss, even though we can do so only in a very preliminary fashion, what relationship there may be between the close visual companion to HD~160934 and the unseen spectroscopic secondary that is responsible for the velocity variations of the primary, whose orbit is gradually maturing.

\begin{table}[h]\begin{center}
\setlength{\doublerulesep}{\arrayrulewidth}
\caption{Tentative orbital elements for HD~160934 \label{table-15}\vspace{2mm}}
\begin{tabular}{@{}ll@{}}
\hline\hline\\[-5pt]
\hskip1.5em$P$      &=~ 4000~$\pm$~330 days \\
\hskip1.5em$T_{}$ &=~ MJD 52452~$\pm$~11 \\
\hskip1.5em$\gamma$ &=~ $-$29.25~$\pm$~0.16 km\,s$^{-1}$ \\
\hskip1.5em$K$  &=~ 7.39 ~$\pm$~0.22 km\,s$^{-1}$  \\
\hskip1.5em$e$      &=~ 0.697 ~$\pm$~0.26            \\
\hskip1.5em$\omega$ &=~ 220.4 ~$\pm$~3.2 degrees         \\
\hskip1.5em$a_{1}$\,sin\,$i$ &=~ 292~$\pm$~28 Gm \\ 
\hskip1.5em$f(m)$ &=~ 0.062~$\pm$~0.010 $M_{\odot}$ \\[5pt]
\multicolumn{2}{l}{R.m.s. residual (wt. 1) = 0.78 km\,s$^{-1}$} \\[3pt]
\hline
\end{tabular}
\end{center}\end{table}

First we establish the identity of the companions found by the two techniques, by comparing the spectroscopic and 'visual' separations. The $a_{1}$\,sin\,$i$ value of the spectroscopic orbit is close to 2 AU, with a current uncertainty of about 10\%. If the companion has a mass that is smaller in the ratio of 1 to 1.21, as Hormuth et al. propose, then the projected separation of the stars will be 2.21 times $a_{1}$\,sin\,$i$, or abut 4.4 AU. At the distance of about $25_{-6}^{+10}$ pc (unusually poorly determined by \textit{Hipparcos}), that projected separation would subtend from 0\arcsec.13 to 0\arcsec.23---just the sort of distance actually observed, leaving no reasonable doubt that the `visual' and spectroscopic companions are one and the same. It should be pointed out that the preliminary orbit that we give has a high eccentricity, about 0.7, so the actual separation of the stars at apastron, which is fairly near where they were observed by \citeauthor{hormuth}, is about 1.7 times the mean that we have just calculated. But it is still in the same range as the observed angular separation, and all the distances are subject to presently unknown projection factors, which are in altogether different planes in the spectroscopic and angular cases, so no more than a very general correspondence is to be expected.

Having assured ourselves that the directly observed visual companion is indeed the star that is responsible for the radial-velocity variations of the primary, we can consider the results of both observational techniques in a preliminary discussion of the three-dimensional orbit. The spectroscopically estimated orbital period is $11\,\pm\,1$ years, so it must be supposed that rather less than one cycle elapsed between the two `visual' observations. The later one occurred at an orbital phase of about 0.37 (0.13 cycles before apastron), so the earlier (HST) one was probably about the same interval \textit{after} the preceding apastron passage. It seems altogether practicable to engineer an orbit such that the relative path of the secondary on the sky passes through the two positions, one before and the other after apastron; it could be expected that the HST position will have been regained by about the time that we are writing now. The orbit would need to be `direct' (anti-clockwise on the sky; position angles increasing) and probably of rather high inclination.

\nocite{*}
\bibliographystyle{spr-mp-nameyear-cnd}
\bibliography{griffin.filizak}

\end{document}